%% file: graphyti.tex
\documentclass[10pt,journal,compsoc]{IEEEtran}
\usepackage{listings}
\usepackage{gnuplot-lua-tikz}
\usepackage{amssymb}
\usepackage{subcaption}
\usepackage{algorithm}
\usepackage{algpseudocode}

\usepackage{hyperref}

\usepackage{enumitem}
\usepackage{authblk}
\usepackage{cite}
\usepackage{wrapfig}

\definecolor{lbcolor}{rgb}{0.9,0.9,0.9}

\newcommand{\graphyti}{Graphyti}

\algblock{ParFor}{EndParFor}
\algnewcommand\algorithmicparfor{\textbf{parfor}}
\algnewcommand\algorithmicpardo{\textbf{do}}
\algnewcommand\algorithmicendparfor{\textbf{end\ parfor}}
\algrenewtext{ParFor}[1]{\algorithmicparfor\ #1\ \algorithmicpardo}
\algrenewtext{EndParFor}{\algorithmicendparfor}
\algrenewcommand\algorithmicindent{1.0em}
\let\oldReturn\Return
\renewcommand{\Return}{\State\oldReturn}

\begin{document}

\title{\graphyti{}: A Semi-External Memory
        Graph Library for FlashGraph}
\author[1]{\rm Disa Mhembere}
\author[1]{\rm Da Zheng}
\author[3]{\rm Carey E. Priebe}
\author[3]{\rm Joshua T. Vogelstein}
\author[1]{\rm Randal Burns}
\affil[1]{Department of Computer Science, Johns Hopkins University}
\affil[3]{Department of Applied Mathematics and Statistics, Johns Hopkins University}
\affil[4]{Department of Biomedical Engineering, Johns Hopkins University}

\IEEEtitleabstractindextext{
\input{abstract}

\begin{IEEEkeywords}
    graph analysis, semi-external memory
\end{IEEEkeywords}}

\maketitle

\IEEEpubid{\begin{minipage}{\textwidth}\ \\[12pt] \centering
\copyright 2019 IEEE. Personal use of this material is permitted.
    Permission from IEEE must be obtained for all other uses, in any current or
    future media, including reprinting/republishing this material for
    advertising or promotional purposes, creating new collective works,
    for resale or redistribution to servers or lists, or reuse of any
    copyrighted component of this work in other works.
\end{minipage}}

\IEEEdisplaynontitleabstractindextext
\IEEEpeerreviewmaketitle

\input{intro}
\input{relwork}
\input{architecture}

\input{principles}

\input{software}
\input{conclusions}

\section*{Acknowledgments}
This work is supported by DARPA GRAPHS N66001-14-1-4028,
DARPA SIMPLEX N66001-15-C-4041, and by NSF ACI-1649880.

\bibliographystyle{abbrv}
\bibliography{graphyti}

\input{author-bios}

\end{document}

%% file: abstract.tex
\begin{abstract}
Graph datasets exceed the in-memory capacity of most
standalone machines. Traditionally, graph frameworks have overcome memory limitations through
scale-out, distributing computing.
Emerging frameworks avoid the network bottleneck of distributed data with
Semi-External Memory (SEM) that uses a single multicore node and operates on
graphs larger than memory.
In SEM, $\mathcal{O}(m)$ data resides on disk and $\mathcal{O}(n)$ data
in memory, for a graph with $n$ vertices and $m$ edges.
For developers, this adds complexity because they must explicitly
encode I/O within applications.
We present principles that are critical for application developers
to adopt in order to achieve state-of-the-art performance,
while minimizing I/O and memory for algorithms in SEM.
We present them in \graphyti{}, an extensible
parallel SEM graph library built on FlashGraph and available in Python via
\texttt{pip}.
In SEM, \graphyti{} achieves 80\% of the performance of in-memory execution and
retains the performance of FlashGraph, which
outperforms distributed engines, such as PowerGraph and Galois.
\end{abstract}

%% file: intro.tex
\IEEEraisesectionheading{\section{Introduction}\label{sec:intro}}

\IEEEPARstart{G}{raphs} datasets exceed the in-memory capacity of modern computers.
With the advent of multi-core NUMA machines and fast solid state storage,
such as NVMe SSDs, developers embraced semi-external memory (SEM) for graph
analytics \cite{flashgraph,graphene,gstore,clip,hpgraph}
to scale algorithms to problems that exceed memory capacity.
Despite their promise, SEM graph engines have experienced relatively low adoption rates,
owing to (i) the challenges of programming SEM algorithms, and
(ii) limited algorithms in SEM graph libraries.

We address fundamental questions regarding efficient and scalable design of
semi-external memory applications within a vertex-centric graph framework.
Vertex-centric programs encode algorithms as actions on the vertexes of a graph
and achieve parallelism by processing many vertices concurrently.
Vertices are \textit{activated} in a bulk synchronous processing (BSP) step that
completes when all vertices are \textit{inactive} and the framework reaches a
global synchronization barrier.
We identify core principles through optimizations within \graphyti{} and
demonstrate improved runtime performance.
\graphyti{} provides a high-level language interface to a broad range of
popular graph algorithms.

Semi-external memory graph frameworks are an attractive alternative to
distributed frameworks, because they avoid the performance penalty from
moving data across networks.
SEM frameworks \cite{flashgraph,graphene,gstore} deliver large-scale graph
processing on limited hardware;
they typically exceed the performance of systems with an order of magnitude
more processing \cite{powergraph,turi}.
As such, understanding how to
achieve highly parallel, I/O-minimal applications is critical to SEM adoption.
This work describes several popular algorithms in SEM and uses these
algorithms as examples of patterns for development of other large-scale graph
applications.

The challenge with developing SEM vertex-centric applications is that
programmers must explicitly request edge data from disk and maintain
$\mathcal{O}(n)$ in-memory state.
SEM adds a layer of complexity as developers must now also
encode I/O and memory usage.
We highlight principles that ease this process for
a wide variety of algorithms.

We present principles and techniques that lower the barrier of entry for
the vertex-centric SEM applications. We illustrate principles through
example applications implemented in an open-source, extensible library---\graphyti{}.
Based on these optimizations, we realize a semi-external memory tool that
realizes 80\% of the performance of totally in-memory computation, reducing
memory consumption by a factor of 20 to 100 of the total graph size.

%% file: relwork.tex
\section{Related Works}

Popular graph libraries \cite{igraph, networkx} are flexible, but
lack multithreaded support and, thus, scalability.
Application development is simple because graph algorithms are invoked
as functions against data.
Performance optimizations revolve around data structure design.
Developers may assume all vertices are in-memory and need not
consider I/O.

Distributed frameworks, such as Turi \cite{turi} and Mahout \cite{mahout},
scale by processing
graphs on multiple nodes across a network.
Datasets must fit in the aggregate memory of a
cluster. Such frameworks encode parallelism through \textit{vertex-centric} or
\textit{edge-centric} computing abstractions.
Google's Pregel \cite{pregel} introduced vertex-centric programming,
which has become the prominent abstraction for graph parallelism
\cite{turi,ligra,galois}.
Network traffic bottlenecks distributed graph frameworks.
Therefore, most optimizations focus on reducing network I/O and do not focus
on reducing memory consumption.
Distributed frameworks use process-level concurrency and do not take advantage
of shared-memory at computing nodes.

Out-of-core graph frameworks \cite{gridgraph,graphchi,xstream}
minimize memory and maximize scalability by streaming datasets from disk.
This comes at a steep performance cost, because each iteration rereads
the entire dataset.
In contrast, SEM holds
$\mathcal{O}(n)$ data in-memory and performs selective I/O to only the edge data
that are needed.

Other out-of-core frameworks rely on heavy graph format preprocessing and
use custom hardware co-processors \cite{mosaic} or GPUs \cite{uberflow}
to improve performance.
Libraries in this space must minimize I/O between device and host.
The architecture, memory hierarchy and processor
density of co-processors differ vastly from that of CPUs. This leads to
programming patterns that are distinct from those that accelerate SEM
applications on CPUs.

SEM realizes high-performance on a modest amount of commodity hardware
and have become the focus of much recent effort
\cite{flashgraph,graphene,gstore,clip,hpgraph, graphmp}.
The SEM model offers a rich optimization landscape that includes I/O reduction,
caching optimization,
I/O prefetching and overlap of computation with asynchronous I/O.
The key difference in programming for SEM is that vertexes must explicitly
issue I/O requests for edge data.
Once requests are fulfilled and data are in memory, \textit{activated} vertices
are processed. Although \graphyti{} is built on FlashGraph,
application optimizations are generic to the SEM model and, thus, could be implemented in
other SEM frameworks.

%% file: architecture.tex
\section{Architecture} \label{sec:gt:arch}

\begin{figure}[t]
    \centering
    \includegraphics[width=\linewidth]{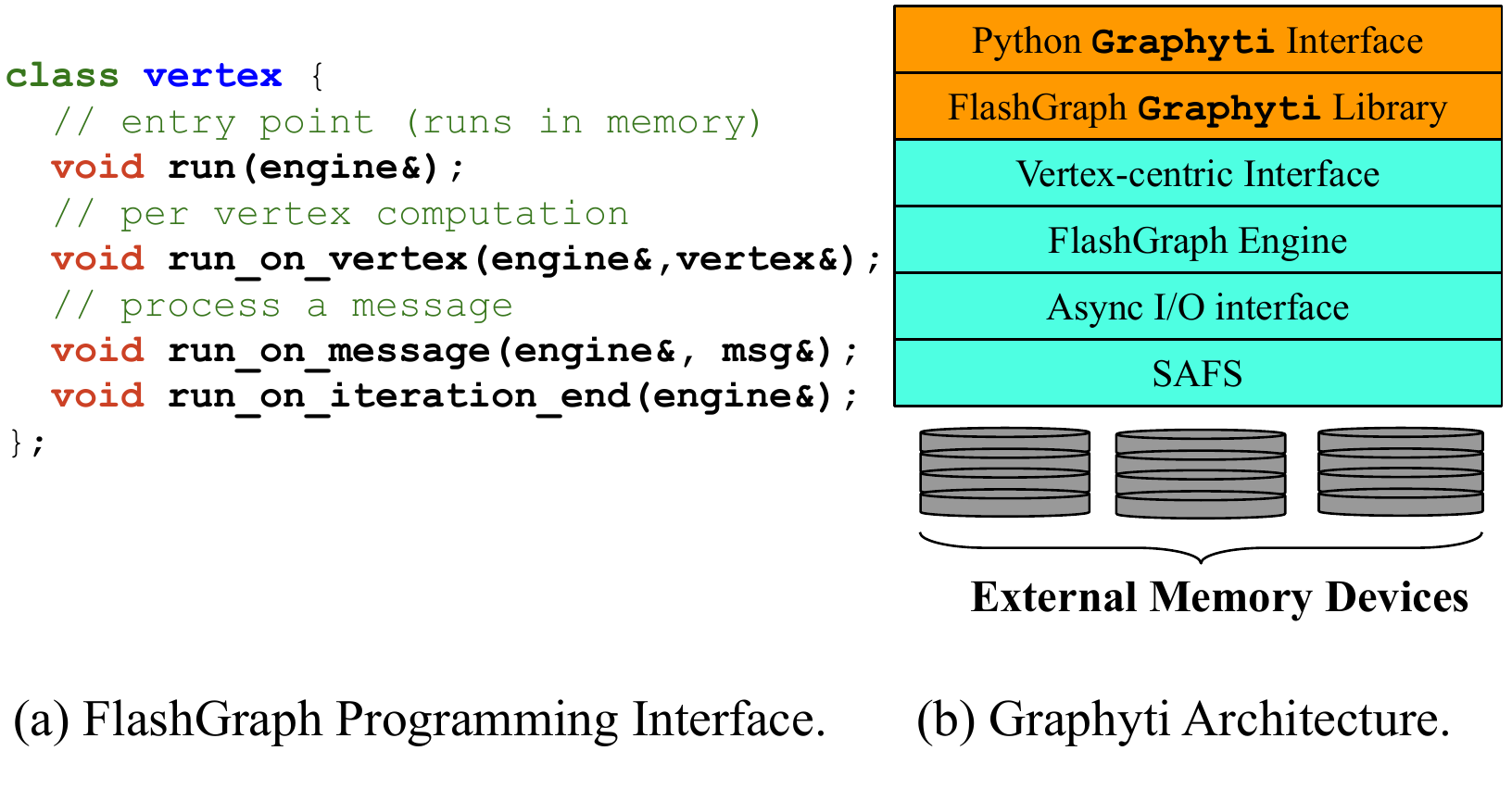}

    \vspace{-10pt}

    \caption{\graphyti{} and FlashGraph.}
    \vspace{-10pt}

    \label{fig:graphyti}
\end{figure}

\graphyti{} provides python bindings and a C++ library that runs on the
FlashGraph engine.
FlashGraph builds upon the SAFS userspace file system \cite{safs} that
performs asynchronous parallel I/O from external memory devices.
SAFS is distributed and installed transparently with FlashGraph.
Figure \ref{fig:graphyti} shows the C++ FlashGraph programming interface and
architecture.

%% file: principles.tex
\section{Principles} \label{sec:gt:principles}

We present six algorithms that demonstrate the principles
that are critical to realize state-of-the-art performance for SEM vertex-centric
applications. The patterns in these algorithms serve as a blueprint for the
developers of other SEM algorithms.
Each subsection (\ref{subsec:gt:pagerank} -- \ref{subsec:gt:louvain})
describes an algorithm followed by the vertex-centric,
SEM optimizations.

We conduct validation experiments on either the directed or undirected version
of the Twitter \cite{twitter-graph} graph
dataset which contains $42$ Million vertices and $1.5$ Billion edges of size
$14$ GB. All experiments require no more than $4$ GB of memory of which
$2$ GB is used for FlashGraph's configurable page cache.

\input{pagerank}
\input{coreness}
\input{diameter}

\input{bc}

\input{tc}
\input{louvain}

%% file: pagerank.tex
\subsection{PageRank} \label{subsec:gt:pagerank}

PageRank \cite{pagerank} is an iterative algorithm that identifies vertices
of high importance in a directed graph. It assigns a higher rank to
vertices referenced by other high ranking vertices as follows:

\vspace{-10pt}
\begin{equation} \label{eqn:pagerank}
R(u) = c \sum_{v \in B_u} \frac{R(v)}{N_v},
\end{equation}
\vspace{-10pt}

\noindent in which $R(x)$ is the PageRank of vertex $x$, $B_x$ is
the set of all inward
pointing neighbors of vertex $x$, $c$ is a normalization factor,
and $N_x$ is the number of outward pointing neighbors of vertex $x$.
Traditionally, developers adopt the following algorithm for vertex-centric
interfaces:

\begin{enumerate}
    \item gather in-edge neighbor PageRank values.
    \item compute a vertex's updated PageRank.
    \item if the updated PageRank value surpasses a predefined threshold,
        multicast to out-bound neighbors informing them to activate.
\end{enumerate}

We refer to this as the \texttt{PR-pull} algorithm and it is utilized by both
Google's Pregel \cite{pregel} and Apple's Turi \cite{turi}. In the
\textit{pull} model vertices extract information from their neighbors.

\begin{figure}[!htb]
    \centering
        \input{./charts/pr}
        \vspace{-10pt}
    \caption[\graphyti{} pagerank performance.]{
        Runtime, Read I/O, I/O requests,
    and thread context switches of \texttt{PR-push} when compared with
    \texttt{PR-pull}.}
    \label{fig:pr}
\end{figure}
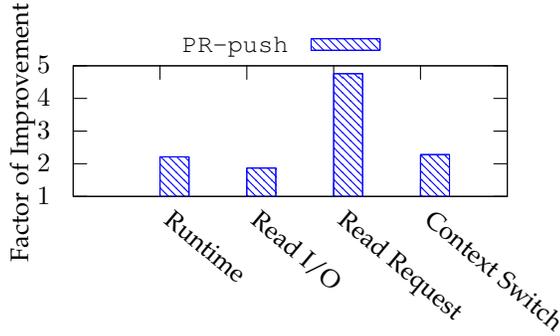

When developing the application for SEM we must prioritize I/O minimization.
We instead adopt a \textit{push} (\texttt{PR-push}) model as follows:

\begin{enumerate}
    \item compute a vertex's PageRank.
    \item if a vertex's current PageRank exceeds a predefined threshold,
        multicast its PageRank to its out-bound neighbors.
\end{enumerate}
\texttt{PR-push} demonstrates the principle:

\textbf{Limit superfluous reads}:
The key insight is that \texttt{PR-pull} activates vertices and
requests data for neighbors whose PageRank has already converged.
\texttt{PR-push} instead computes a delta then sends messages only
activating the minimal subset of vertices necessary.
Vertex activation, processing and the superfluous I/O degrade
the performance of \texttt{PR-pull}.
Even though \texttt{PR-push} and \texttt{PR-pull} share the same upper bound of
messaging complexity ($\mathcal{O}(m^2)$), \texttt{PR-push} sends
fewer messages, reducing I/O and improving performance.
Figure \ref{fig:pr} demonstrates a reduction of I/O
by a factor of $1.8$, and improvement in runtime of $2.2$.
Furthermore, \texttt{PR-push} reduces I/O read requests by a factor of nearly $5$.
Finally, a reduction in messages leads to reduced burden on
FlashGraph to load balance message queues for worker threads.

%% file: charts/pr.tex
\begin{tikzpicture}[gnuplot]
\path (0.000,0.000) rectangle (7.620,4.318);
\gpcolor{color=gp lt color border}
\gpsetlinetype{gp lt border}
\gpsetlinewidth{1.00}
\draw[gp path] (0.952,1.963)--(1.132,1.963);
\draw[gp path] (6.723,1.963)--(6.543,1.963);
\node[gp node right] at (0.768,1.963) {$1$};
\draw[gp path] (0.952,2.398)--(1.132,2.398);
\draw[gp path] (6.723,2.398)--(6.543,2.398);
\node[gp node right] at (0.768,2.398) {$2$};
\draw[gp path] (0.952,2.832)--(1.132,2.832);
\draw[gp path] (6.723,2.832)--(6.543,2.832);
\node[gp node right] at (0.768,2.832) {$3$};
\draw[gp path] (0.952,3.267)--(1.132,3.267);
\draw[gp path] (6.723,3.267)--(6.543,3.267);
\node[gp node right] at (0.768,3.267) {$4$};
\draw[gp path] (0.952,3.701)--(1.132,3.701);
\draw[gp path] (6.723,3.701)--(6.543,3.701);
\node[gp node right] at (0.768,3.701) {$5$};
\draw[gp path] (2.106,1.963)--(2.106,2.143);
\draw[gp path] (2.106,3.701)--(2.106,3.521);
\node[gp node left,rotate=-40] at (2.106,1.779) {Runtime};
\draw[gp path] (3.260,1.963)--(3.260,2.143);
\draw[gp path] (3.260,3.701)--(3.260,3.521);
\node[gp node left,rotate=-40] at (3.260,1.779) {Read I/O};
\draw[gp path] (4.415,1.963)--(4.415,2.143);
\draw[gp path] (4.415,3.701)--(4.415,3.521);
\node[gp node left,rotate=-40] at (4.415,1.779) {Read Request};
\draw[gp path] (5.569,1.963)--(5.569,2.143);
\draw[gp path] (5.569,3.701)--(5.569,3.521);
\node[gp node left,rotate=-40] at (5.569,1.779) {Context Switch};
\draw[gp path] (0.952,3.701)--(0.952,1.963)--(6.723,1.963)--(6.723,3.701)--cycle;
\node[gp node center,rotate=-270] at (0.276,2.832) {Factor of Improvement};
\node[gp node right] at (3.931,3.984) {\texttt{PR-push}};
\def\gpfillpath{(4.115,3.907)--(5.031,3.907)--(5.031,4.061)--(4.115,4.061)--cycle}
\gpfill{color=gpbgfillcolor} \gpfillpath;
\gpfill{rgb color={0.000,0.000,1.000},gp pattern 2,pattern color=.} \gpfillpath;
\gpcolor{rgb color={0.000,0.000,1.000}}
\draw[gp path] (4.115,3.907)--(5.031,3.907)--(5.031,4.061)--(4.115,4.061)--cycle;
\def\gpfillpath{(2.106,1.963)--(2.492,1.963)--(2.492,2.490)--(2.106,2.490)--cycle}
\gpfill{color=gpbgfillcolor} \gpfillpath;
\gpfill{rgb color={0.000,0.000,1.000},gp pattern 2,pattern color=.} \gpfillpath;
\draw[gp path] (2.106,1.963)--(2.106,2.489)--(2.491,2.489)--(2.491,1.963)--cycle;
\def\gpfillpath{(3.260,1.963)--(3.646,1.963)--(3.646,2.341)--(3.260,2.341)--cycle}
\gpfill{color=gpbgfillcolor} \gpfillpath;
\gpfill{rgb color={0.000,0.000,1.000},gp pattern 2,pattern color=.} \gpfillpath;
\draw[gp path] (3.260,1.963)--(3.260,2.340)--(3.645,2.340)--(3.645,1.963)--cycle;
\def\gpfillpath{(4.415,1.963)--(4.800,1.963)--(4.800,3.596)--(4.415,3.596)--cycle}
\gpfill{color=gpbgfillcolor} \gpfillpath;
\gpfill{rgb color={0.000,0.000,1.000},gp pattern 2,pattern color=.} \gpfillpath;
\draw[gp path] (4.415,1.963)--(4.415,3.595)--(4.799,3.595)--(4.799,1.963)--cycle;
\def\gpfillpath{(5.569,1.963)--(5.955,1.963)--(5.955,2.520)--(5.569,2.520)--cycle}
\gpfill{color=gpbgfillcolor} \gpfillpath;
\gpfill{rgb color={0.000,0.000,1.000},gp pattern 2,pattern color=.} \gpfillpath;
\draw[gp path] (5.569,1.963)--(5.569,2.519)--(5.954,2.519)--(5.954,1.963)--cycle;
\gpcolor{color=gp lt color border}
\draw[gp path] (0.952,3.701)--(0.952,1.963)--(6.723,1.963)--(6.723,3.701)--cycle;
\gpdefrectangularnode{gp plot 1}{\pgfpoint{0.952cm}{1.963cm}}{\pgfpoint{6.723cm}{3.701cm}}
\end{tikzpicture}

%% file: coreness.tex
\subsection{Coreness Decomposition} \label{subsec:gt:coreness}

\begin{figure}[!htb]
\centering
        \input{./charts/coreness}
    \vspace{-10pt}
    \caption[\graphyti{} coreness runtime analysis.]{
        Performance improvement of
    \graphyti{}'s \texttt{coreness} compared to an implementation
    with point-to-point messages and no pruning. Pruning + hybrid messaging
    is $2.3$X faster than pruning alone and $60$X faster than unoptimized.}
\label{fig:coreness}
\end{figure}
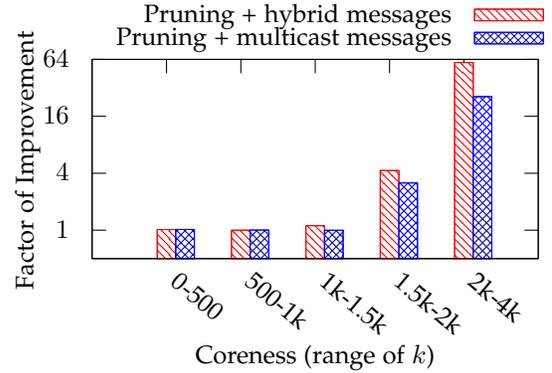

Coreness decomposition extracts a maximal subgraph in which each vertex
has at least degree $k_{max}$. The algorithm proceeds by
iteratively deleting vertices beginning with those with degree $0$
until $k_{max}$.
Deleted vertices notify neighboring vertices to reduce their degree until
only vertices with a coreness of $\geq k_{max}$ remain. The optimizations we
employ to improve the performance of coreness highlight the following core
principles:

\textbf{Minimize messaging}:
\graphyti{}'s \texttt{coreness} adopts a hybrid messaging discipline inspired by
guided schedulers. In early iterations, almost all vertices modify
their degree and inform neighbors of edge and node deletions.
During this phase, multicast messages are most efficient.
As the graph becomes sparser, multicast messages incur higher overhead because
many neighboring vertices with lower coreness values have already been deleted.
At this point, point-to-point messages greatly reduces messaging overhead,
improving runtime as shown in Figure \ref{fig:coreness}. \graphyti{}'s
coreness maintains a distribution over all remaining vertices to
determine when each one should should switch to point-to-point messaging.
We empirically determine that once a vertex has $10\%$ of its original degree,
point-to-point messaging improves the time necessary to process a single vertex
by an order of magnitude.

\textbf{Algorithmically prune computation}:
At the completion of a coreness iteration, $k_i$, in which $k_i < k_{max}$, as
stated, the algorithm would proceed to $k_{i+1}$,
$k_{i+2}$ and so forth. \graphyti{} prunes unnecessary $k_i$ values by observing the
next possible core value is at least $k_{\min(deg(\alpha))} \forall \alpha \in A$,
in which $deg$ the degree of a vertex, $\alpha \subset V$ and $V$ is the set of
all vertices in the graph. This optimization alone improves performance by
an order of magnitude (Figure \ref{fig:coreness}).


%% file: charts/coreness.tex
\begin{tikzpicture}[gnuplot]
\path (0.000,0.000) rectangle (7.620,5.080);
\gpcolor{color=gp lt color border}
\gpsetlinetype{gp lt border}
\gpsetlinewidth{1.00}
\draw[gp path] (1.136,1.883)--(1.316,1.883);
\draw[gp path] (7.067,1.883)--(6.887,1.883);
\node[gp node right] at (0.952,1.883) {$1$};
\draw[gp path] (1.136,2.640)--(1.316,2.640);
\draw[gp path] (7.067,2.640)--(6.887,2.640);
\node[gp node right] at (0.952,2.640) {$4$};
\draw[gp path] (1.136,3.398)--(1.316,3.398);
\draw[gp path] (7.067,3.398)--(6.887,3.398);
\node[gp node right] at (0.952,3.398) {$16$};
\draw[gp path] (1.136,4.155)--(1.316,4.155);
\draw[gp path] (7.067,4.155)--(6.887,4.155);
\node[gp node right] at (0.952,4.155) {$64$};
\draw[gp path] (2.125,1.504)--(2.125,1.684);
\draw[gp path] (2.125,4.155)--(2.125,3.975);
\node[gp node left,rotate=-40] at (2.125,1.320) {0-500};
\draw[gp path] (3.113,1.504)--(3.113,1.684);
\draw[gp path] (3.113,4.155)--(3.113,3.975);
\node[gp node left,rotate=-40] at (3.113,1.320) {500-1k};
\draw[gp path] (4.102,1.504)--(4.102,1.684);
\draw[gp path] (4.102,4.155)--(4.102,3.975);
\node[gp node left,rotate=-40] at (4.102,1.320) {1k-1.5k};
\draw[gp path] (5.090,1.504)--(5.090,1.684);
\draw[gp path] (5.090,4.155)--(5.090,3.975);
\node[gp node left,rotate=-40] at (5.090,1.320) {1.5k-2k};
\draw[gp path] (6.079,1.504)--(6.079,1.684);
\draw[gp path] (6.079,4.155)--(6.079,3.975);
\node[gp node left,rotate=-40] at (6.079,1.320) {2k-4k};
\draw[gp path] (1.136,4.155)--(1.136,1.504)--(7.067,1.504)--(7.067,4.155)--cycle;
\node[gp node center,rotate=-270] at (0.276,2.829) {Factor of Improvement};
\node[gp node center] at (4.101,0.215) {Coreness (range of $k$)};
\node[gp node right] at (6.035,4.746) {Pruning + hybrid messages};
\def\gpfillpath{(6.219,4.669)--(7.135,4.669)--(7.135,4.823)--(6.219,4.823)--cycle}
\gpfill{color=gpbgfillcolor} \gpfillpath;
\gpfill{rgb color={1.000,0.000,0.000},gp pattern 2,pattern color=.} \gpfillpath;
\gpcolor{rgb color={1.000,0.000,0.000}}
\draw[gp path] (6.219,4.669)--(7.135,4.669)--(7.135,4.823)--(6.219,4.823)--cycle;
\def\gpfillpath{(2.001,1.504)--(2.249,1.504)--(2.249,1.892)--(2.001,1.892)--cycle}
\gpfill{color=gpbgfillcolor} \gpfillpath;
\gpfill{rgb color={1.000,0.000,0.000},gp pattern 2,pattern color=.} \gpfillpath;
\draw[gp path] (2.001,1.504)--(2.001,1.891)--(2.248,1.891)--(2.248,1.504)--cycle;
\def\gpfillpath{(2.989,1.504)--(3.238,1.504)--(3.238,1.883)--(2.989,1.883)--cycle}
\gpfill{color=gpbgfillcolor} \gpfillpath;
\gpfill{rgb color={1.000,0.000,0.000},gp pattern 2,pattern color=.} \gpfillpath;
\draw[gp path] (2.989,1.504)--(2.989,1.882)--(3.237,1.882)--(3.237,1.504)--cycle;
\def\gpfillpath{(3.978,1.504)--(4.226,1.504)--(4.226,1.943)--(3.978,1.943)--cycle}
\gpfill{color=gpbgfillcolor} \gpfillpath;
\gpfill{rgb color={1.000,0.000,0.000},gp pattern 2,pattern color=.} \gpfillpath;
\draw[gp path] (3.978,1.504)--(3.978,1.942)--(4.225,1.942)--(4.225,1.504)--cycle;
\def\gpfillpath{(4.966,1.504)--(5.215,1.504)--(5.215,2.677)--(4.966,2.677)--cycle}
\gpfill{color=gpbgfillcolor} \gpfillpath;
\gpfill{rgb color={1.000,0.000,0.000},gp pattern 2,pattern color=.} \gpfillpath;
\draw[gp path] (4.966,1.504)--(4.966,2.676)--(5.214,2.676)--(5.214,1.504)--cycle;
\def\gpfillpath{(5.955,1.504)--(6.203,1.504)--(6.203,4.114)--(5.955,4.114)--cycle}
\gpfill{color=gpbgfillcolor} \gpfillpath;
\gpfill{rgb color={1.000,0.000,0.000},gp pattern 2,pattern color=.} \gpfillpath;
\draw[gp path] (5.955,1.504)--(5.955,4.113)--(6.202,4.113)--(6.202,1.504)--cycle;
\gpcolor{color=gp lt color border}
\node[gp node right] at (6.035,4.438) {Pruning + multicast messages};
\def\gpfillpath{(6.219,4.361)--(7.135,4.361)--(7.135,4.515)--(6.219,4.515)--cycle}
\gpfill{color=gpbgfillcolor} \gpfillpath;
\gpfill{rgb color={0.000,0.000,1.000},gp pattern 3,pattern color=.} \gpfillpath;
\gpcolor{rgb color={0.000,0.000,1.000}}
\draw[gp path] (6.219,4.361)--(7.135,4.361)--(7.135,4.515)--(6.219,4.515)--cycle;
\def\gpfillpath{(2.248,1.504)--(2.496,1.504)--(2.496,1.893)--(2.248,1.893)--cycle}
\gpfill{color=gpbgfillcolor} \gpfillpath;
\gpfill{rgb color={0.000,0.000,1.000},gp pattern 3,pattern color=.} \gpfillpath;
\draw[gp path] (2.248,1.504)--(2.248,1.892)--(2.495,1.892)--(2.495,1.504)--cycle;
\def\gpfillpath{(3.237,1.504)--(3.485,1.504)--(3.485,1.885)--(3.237,1.885)--cycle}
\gpfill{color=gpbgfillcolor} \gpfillpath;
\gpfill{rgb color={0.000,0.000,1.000},gp pattern 3,pattern color=.} \gpfillpath;
\draw[gp path] (3.237,1.504)--(3.237,1.884)--(3.484,1.884)--(3.484,1.504)--cycle;
\def\gpfillpath{(4.225,1.504)--(4.473,1.504)--(4.473,1.882)--(4.225,1.882)--cycle}
\gpfill{color=gpbgfillcolor} \gpfillpath;
\gpfill{rgb color={0.000,0.000,1.000},gp pattern 3,pattern color=.} \gpfillpath;
\draw[gp path] (4.225,1.504)--(4.225,1.881)--(4.472,1.881)--(4.472,1.504)--cycle;
\def\gpfillpath{(5.214,1.504)--(5.462,1.504)--(5.462,2.512)--(5.214,2.512)--cycle}
\gpfill{color=gpbgfillcolor} \gpfillpath;
\gpfill{rgb color={0.000,0.000,1.000},gp pattern 3,pattern color=.} \gpfillpath;
\draw[gp path] (5.214,1.504)--(5.214,2.511)--(5.461,2.511)--(5.461,1.504)--cycle;
\def\gpfillpath{(6.202,1.504)--(6.450,1.504)--(6.450,3.660)--(6.202,3.660)--cycle}
\gpfill{color=gpbgfillcolor} \gpfillpath;
\gpfill{rgb color={0.000,0.000,1.000},gp pattern 3,pattern color=.} \gpfillpath;
\draw[gp path] (6.202,1.504)--(6.202,3.659)--(6.449,3.659)--(6.449,1.504)--cycle;
\gpcolor{color=gp lt color border}
\draw[gp path] (1.136,4.155)--(1.136,1.504)--(7.067,1.504)--(7.067,4.155)--cycle;
\gpdefrectangularnode{gp plot 1}{\pgfpoint{1.136cm}{1.504cm}}{\pgfpoint{7.067cm}{4.155cm}}
\end{tikzpicture}

%% file: diameter.tex
\subsection{Graph Diameter} \label{subsec:gt:diameter}

Graph diameter for connected graphs is the maximum of the
all pairs shortest paths in a graph. Exact graph diameter has
complexity $\mathcal{O}(n^3)$ and is computationally challenging for
any framework. \graphyti{} computes an estimated diameter using a
series of breadth-first searches from \textit{pseudo-peripheral} vertices:
ones as close to the extremities of the graph as possible. Optimizing diameter estimation
highlights the following guiding principle:

\textbf{Decouple algorithm development from framework constructs}:
Diameter estimation provides the opportunity to design a more efficient vertex-centric
application. A simple algorithm repeats the following until all reachable vertices are visited:

\begin{enumerate}
    \item select a peripheral \textit{source} vertex.
    \item perform BFS from the selected vertex.
    \item update neighboring vertex distances to one greater than
        their nearest neighbor in parallel.
\end{enumerate}

This \textit{uni-source} BFS can be performed multiple times
with different source vertices to find larger diameters.
Although parallel, this algorithm limits the
potential amount of work each vertex performs in a single BFS iteration,
limiting CPU cache data reuse,
and increasing the relative overhead of synchronization barriers at each BSP step.
Uni-source BFS does not reuse edge data that are brought into memory, resulting
in increased data stalls as the application becomes heavily I/O bound.

\begin{figure}[t]
\centering
\includegraphics[width=\linewidth]{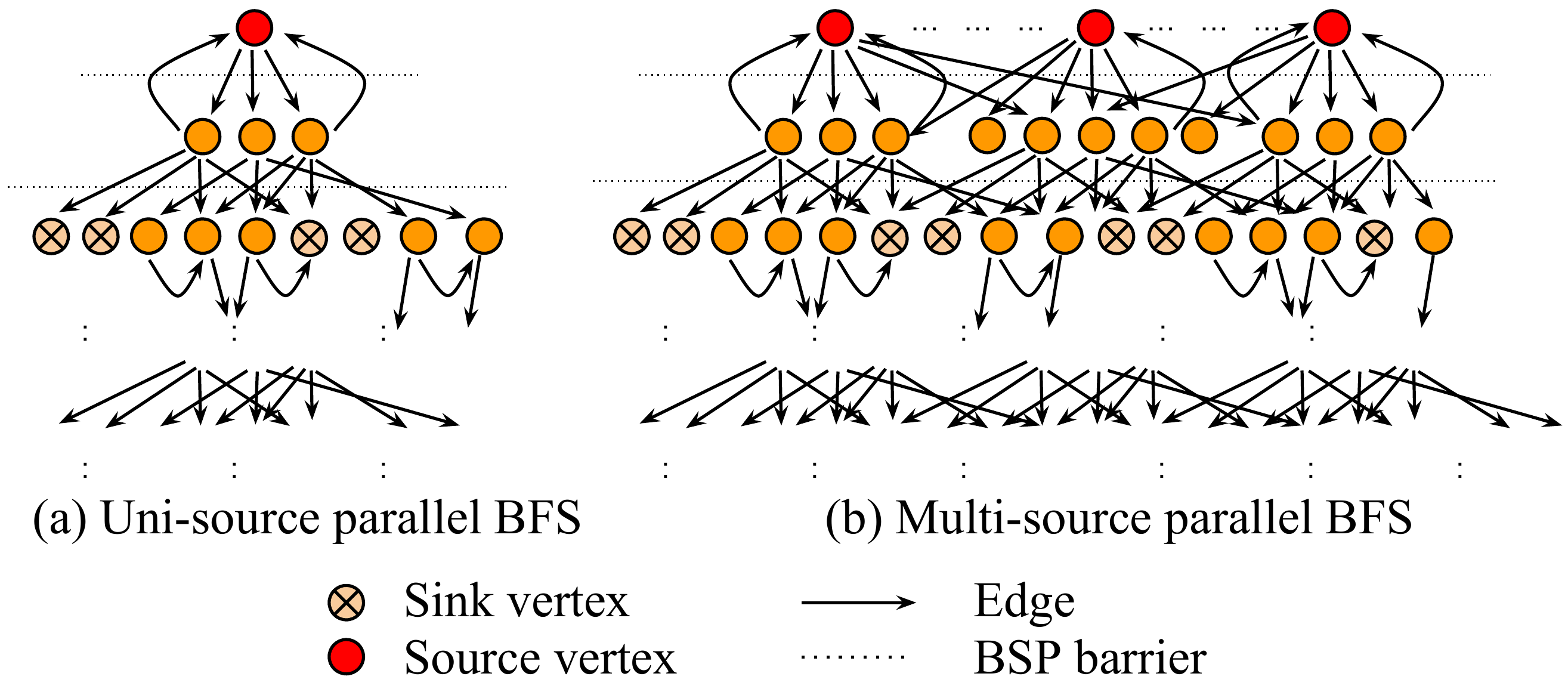}
    \caption[Uni-source vs. Multi-source BFS]{
        Uni-source BFS (left) is susceptible to terminal paths due to sink
    vertices and loops. Multi-source BFS (right) increases page cache hits
    because multiple paths activate the same vertices in each BFS frontier.}
\label{fig:ms-bfs}
\end{figure}

\begin{figure}[!htb]
\centering
\footnotesize
\begin{subfigure}{.4\textwidth}
\centering
    \input{./charts/dia-io}
    \vspace{-10pt}
    \caption{Quantity of data read from SSDs.}
\label{fig:dia-io}
\end{subfigure}

\begin{subfigure}{.4\textwidth}
    \input{./charts/dia-perf}
    \vspace{-10pt}
\caption{Runtime performance.}
\label{fig:dia-perf}
\end{subfigure}

    \caption[\graphyti{} Uni-source vs. Multi-source BFS.]{
        I/O and Runtime comparison of uni-source BFS and parallel
    multi-source BFS in \texttt{diameter}.}
\label{fig:dia}
\end{figure}
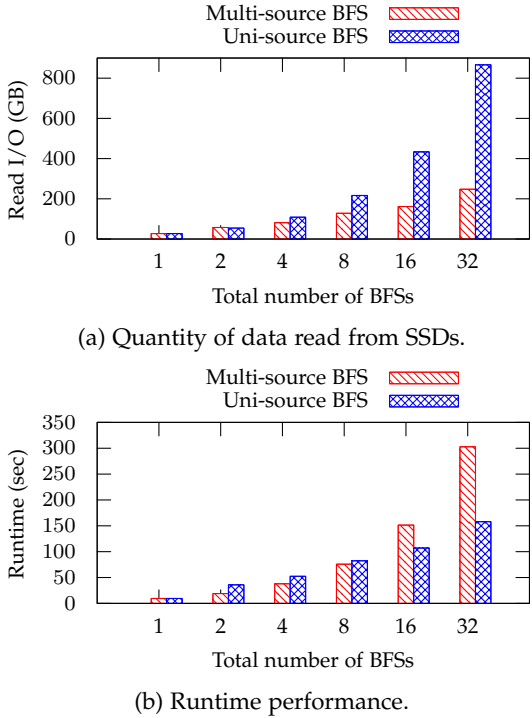

\graphyti{} rethinks the computation to minimize the overhead of
each BSP step by performing concurrent parallel breadth-first
searches (Figure \ref{fig:ms-bfs}).
This \textit{multi-source} BFS leads to higher rates of vertex activations
within a BSP step and consequently lower global barrier overhead compared with
uni-source BFS. Additionally, this reduces cache
thrashing, because requested data that are now in-memory have greater
opportunity for reuse. In multi-source BFS, each vertex
holds a bitmap indicating which BFS path(s) it is on.
Figure \ref{fig:dia-perf} demonstrates the performance
improvements and I/O reduction induced by these optimizations.

%% file: charts/dia-io.tex
\begin{tikzpicture}[gnuplot]
\path (0.000,0.000) rectangle (7.620,4.318);
\gpcolor{color=gp lt color border}
\gpsetlinetype{gp lt border}
\gpsetlinewidth{1.00}
\draw[gp path] (1.320,0.985)--(1.500,0.985);
\draw[gp path] (7.067,0.985)--(6.887,0.985);
\node[gp node right] at (1.136,0.985) {$0$};
\draw[gp path] (1.320,1.520)--(1.500,1.520);
\draw[gp path] (7.067,1.520)--(6.887,1.520);
\node[gp node right] at (1.136,1.520) {$200$};
\draw[gp path] (1.320,2.055)--(1.500,2.055);
\draw[gp path] (7.067,2.055)--(6.887,2.055);
\node[gp node right] at (1.136,2.055) {$400$};
\draw[gp path] (1.320,2.590)--(1.500,2.590);
\draw[gp path] (7.067,2.590)--(6.887,2.590);
\node[gp node right] at (1.136,2.590) {$600$};
\draw[gp path] (1.320,3.125)--(1.500,3.125);
\draw[gp path] (7.067,3.125)--(6.887,3.125);
\node[gp node right] at (1.136,3.125) {$800$};
\draw[gp path] (2.141,0.985)--(2.141,1.165);
\draw[gp path] (2.141,3.393)--(2.141,3.213);
\node[gp node center] at (2.141,0.677) {1};
\draw[gp path] (2.962,0.985)--(2.962,1.165);
\draw[gp path] (2.962,3.393)--(2.962,3.213);
\node[gp node center] at (2.962,0.677) {2};
\draw[gp path] (3.783,0.985)--(3.783,1.165);
\draw[gp path] (3.783,3.393)--(3.783,3.213);
\node[gp node center] at (3.783,0.677) {4};
\draw[gp path] (4.604,0.985)--(4.604,1.165);
\draw[gp path] (4.604,3.393)--(4.604,3.213);
\node[gp node center] at (4.604,0.677) {8};
\draw[gp path] (5.425,0.985)--(5.425,1.165);
\draw[gp path] (5.425,3.393)--(5.425,3.213);
\node[gp node center] at (5.425,0.677) {16};
\draw[gp path] (6.246,0.985)--(6.246,1.165);
\draw[gp path] (6.246,3.393)--(6.246,3.213);
\node[gp node center] at (6.246,0.677) {32};
\draw[gp path] (1.320,3.393)--(1.320,0.985)--(7.067,0.985)--(7.067,3.393)--cycle;
\node[gp node center,rotate=-270] at (0.276,2.189) {Read I/O (GB)};
\node[gp node center] at (4.193,0.215) {Total number of BFSs};
\node[gp node right] at (5.023,3.984) {Multi-source BFS};
\def\gpfillpath{(5.207,3.907)--(6.123,3.907)--(6.123,4.061)--(5.207,4.061)--cycle}
\gpfill{color=gpbgfillcolor} \gpfillpath;
\gpfill{rgb color={1.000,0.000,0.000},gp pattern 2,pattern color=.} \gpfillpath;
\gpcolor{rgb color={1.000,0.000,0.000}}
\draw[gp path] (5.207,3.907)--(6.123,3.907)--(6.123,4.061)--(5.207,4.061)--cycle;
\def\gpfillpath{(2.038,0.985)--(2.245,0.985)--(2.245,1.058)--(2.038,1.058)--cycle}
\gpfill{color=gpbgfillcolor} \gpfillpath;
\gpfill{rgb color={1.000,0.000,0.000},gp pattern 2,pattern color=.} \gpfillpath;
\draw[gp path] (2.038,0.985)--(2.038,1.057)--(2.244,1.057)--(2.244,0.985)--cycle;
\def\gpfillpath{(2.859,0.985)--(3.066,0.985)--(3.066,1.139)--(2.859,1.139)--cycle}
\gpfill{color=gpbgfillcolor} \gpfillpath;
\gpfill{rgb color={1.000,0.000,0.000},gp pattern 2,pattern color=.} \gpfillpath;
\draw[gp path] (2.859,0.985)--(2.859,1.138)--(3.065,1.138)--(3.065,0.985)--cycle;
\def\gpfillpath{(3.680,0.985)--(3.887,0.985)--(3.887,1.203)--(3.680,1.203)--cycle}
\gpfill{color=gpbgfillcolor} \gpfillpath;
\gpfill{rgb color={1.000,0.000,0.000},gp pattern 2,pattern color=.} \gpfillpath;
\draw[gp path] (3.680,0.985)--(3.680,1.202)--(3.886,1.202)--(3.886,0.985)--cycle;
\def\gpfillpath{(4.501,0.985)--(4.708,0.985)--(4.708,1.328)--(4.501,1.328)--cycle}
\gpfill{color=gpbgfillcolor} \gpfillpath;
\gpfill{rgb color={1.000,0.000,0.000},gp pattern 2,pattern color=.} \gpfillpath;
\draw[gp path] (4.501,0.985)--(4.501,1.327)--(4.707,1.327)--(4.707,0.985)--cycle;
\def\gpfillpath{(5.322,0.985)--(5.529,0.985)--(5.529,1.418)--(5.322,1.418)--cycle}
\gpfill{color=gpbgfillcolor} \gpfillpath;
\gpfill{rgb color={1.000,0.000,0.000},gp pattern 2,pattern color=.} \gpfillpath;
\draw[gp path] (5.322,0.985)--(5.322,1.417)--(5.528,1.417)--(5.528,0.985)--cycle;
\def\gpfillpath{(6.143,0.985)--(6.350,0.985)--(6.350,1.648)--(6.143,1.648)--cycle}
\gpfill{color=gpbgfillcolor} \gpfillpath;
\gpfill{rgb color={1.000,0.000,0.000},gp pattern 2,pattern color=.} \gpfillpath;
\draw[gp path] (6.143,0.985)--(6.143,1.647)--(6.349,1.647)--(6.349,0.985)--cycle;
\gpcolor{color=gp lt color border}
\node[gp node right] at (5.023,3.676) {Uni-source BFS};
\def\gpfillpath{(5.207,3.599)--(6.123,3.599)--(6.123,3.753)--(5.207,3.753)--cycle}
\gpfill{color=gpbgfillcolor} \gpfillpath;
\gpfill{rgb color={0.000,0.000,1.000},gp pattern 3,pattern color=.} \gpfillpath;
\gpcolor{rgb color={0.000,0.000,1.000}}
\draw[gp path] (5.207,3.599)--(6.123,3.599)--(6.123,3.753)--(5.207,3.753)--cycle;
\def\gpfillpath{(2.244,0.985)--(2.450,0.985)--(2.450,1.058)--(2.244,1.058)--cycle}
\gpfill{color=gpbgfillcolor} \gpfillpath;
\gpfill{rgb color={0.000,0.000,1.000},gp pattern 3,pattern color=.} \gpfillpath;
\draw[gp path] (2.244,0.985)--(2.244,1.057)--(2.449,1.057)--(2.449,0.985)--cycle;
\def\gpfillpath{(3.065,0.985)--(3.271,0.985)--(3.271,1.131)--(3.065,1.131)--cycle}
\gpfill{color=gpbgfillcolor} \gpfillpath;
\gpfill{rgb color={0.000,0.000,1.000},gp pattern 3,pattern color=.} \gpfillpath;
\draw[gp path] (3.065,0.985)--(3.065,1.130)--(3.270,1.130)--(3.270,0.985)--cycle;
\def\gpfillpath{(3.886,0.985)--(4.092,0.985)--(4.092,1.276)--(3.886,1.276)--cycle}
\gpfill{color=gpbgfillcolor} \gpfillpath;
\gpfill{rgb color={0.000,0.000,1.000},gp pattern 3,pattern color=.} \gpfillpath;
\draw[gp path] (3.886,0.985)--(3.886,1.275)--(4.091,1.275)--(4.091,0.985)--cycle;
\def\gpfillpath{(4.707,0.985)--(4.913,0.985)--(4.913,1.565)--(4.707,1.565)--cycle}
\gpfill{color=gpbgfillcolor} \gpfillpath;
\gpfill{rgb color={0.000,0.000,1.000},gp pattern 3,pattern color=.} \gpfillpath;
\draw[gp path] (4.707,0.985)--(4.707,1.564)--(4.912,1.564)--(4.912,0.985)--cycle;
\def\gpfillpath{(5.528,0.985)--(5.734,0.985)--(5.734,2.145)--(5.528,2.145)--cycle}
\gpfill{color=gpbgfillcolor} \gpfillpath;
\gpfill{rgb color={0.000,0.000,1.000},gp pattern 3,pattern color=.} \gpfillpath;
\draw[gp path] (5.528,0.985)--(5.528,2.144)--(5.733,2.144)--(5.733,0.985)--cycle;
\def\gpfillpath{(6.349,0.985)--(6.555,0.985)--(6.555,3.303)--(6.349,3.303)--cycle}
\gpfill{color=gpbgfillcolor} \gpfillpath;
\gpfill{rgb color={0.000,0.000,1.000},gp pattern 3,pattern color=.} \gpfillpath;
\draw[gp path] (6.349,0.985)--(6.349,3.302)--(6.554,3.302)--(6.554,0.985)--cycle;
\gpcolor{color=gp lt color border}
\draw[gp path] (1.320,3.393)--(1.320,0.985)--(7.067,0.985)--(7.067,3.393)--cycle;
\gpdefrectangularnode{gp plot 1}{\pgfpoint{1.320cm}{0.985cm}}{\pgfpoint{7.067cm}{3.393cm}}
\end{tikzpicture}

%% file: charts/dia-perf.tex
\begin{tikzpicture}[gnuplot]
\path (0.000,0.000) rectangle (7.620,4.318);
\gpcolor{color=gp lt color border}
\gpsetlinetype{gp lt border}
\gpsetlinewidth{1.00}
\draw[gp path] (1.320,0.985)--(1.500,0.985);
\draw[gp path] (7.067,0.985)--(6.887,0.985);
\node[gp node right] at (1.136,0.985) {$0$};
\draw[gp path] (1.320,1.329)--(1.500,1.329);
\draw[gp path] (7.067,1.329)--(6.887,1.329);
\node[gp node right] at (1.136,1.329) {$50$};
\draw[gp path] (1.320,1.673)--(1.500,1.673);
\draw[gp path] (7.067,1.673)--(6.887,1.673);
\node[gp node right] at (1.136,1.673) {$100$};
\draw[gp path] (1.320,2.017)--(1.500,2.017);
\draw[gp path] (7.067,2.017)--(6.887,2.017);
\node[gp node right] at (1.136,2.017) {$150$};
\draw[gp path] (1.320,2.361)--(1.500,2.361);
\draw[gp path] (7.067,2.361)--(6.887,2.361);
\node[gp node right] at (1.136,2.361) {$200$};
\draw[gp path] (1.320,2.705)--(1.500,2.705);
\draw[gp path] (7.067,2.705)--(6.887,2.705);
\node[gp node right] at (1.136,2.705) {$250$};
\draw[gp path] (1.320,3.049)--(1.500,3.049);
\draw[gp path] (7.067,3.049)--(6.887,3.049);
\node[gp node right] at (1.136,3.049) {$300$};
\draw[gp path] (1.320,3.393)--(1.500,3.393);
\draw[gp path] (7.067,3.393)--(6.887,3.393);
\node[gp node right] at (1.136,3.393) {$350$};
\draw[gp path] (2.141,0.985)--(2.141,1.165);
\draw[gp path] (2.141,3.393)--(2.141,3.213);
\node[gp node center] at (2.141,0.677) {1};
\draw[gp path] (2.962,0.985)--(2.962,1.165);
\draw[gp path] (2.962,3.393)--(2.962,3.213);
\node[gp node center] at (2.962,0.677) {2};
\draw[gp path] (3.783,0.985)--(3.783,1.165);
\draw[gp path] (3.783,3.393)--(3.783,3.213);
\node[gp node center] at (3.783,0.677) {4};
\draw[gp path] (4.604,0.985)--(4.604,1.165);
\draw[gp path] (4.604,3.393)--(4.604,3.213);
\node[gp node center] at (4.604,0.677) {8};
\draw[gp path] (5.425,0.985)--(5.425,1.165);
\draw[gp path] (5.425,3.393)--(5.425,3.213);
\node[gp node center] at (5.425,0.677) {16};
\draw[gp path] (6.246,0.985)--(6.246,1.165);
\draw[gp path] (6.246,3.393)--(6.246,3.213);
\node[gp node center] at (6.246,0.677) {32};
\draw[gp path] (1.320,3.393)--(1.320,0.985)--(7.067,0.985)--(7.067,3.393)--cycle;
\node[gp node center,rotate=-270] at (0.276,2.189) {Runtime (sec)};
\node[gp node center] at (4.193,0.215) {Total number of BFSs};
\node[gp node right] at (5.023,3.984) {Multi-source BFS};
\def\gpfillpath{(5.207,3.907)--(6.123,3.907)--(6.123,4.061)--(5.207,4.061)--cycle}
\gpfill{color=gpbgfillcolor} \gpfillpath;
\gpfill{rgb color={1.000,0.000,0.000},gp pattern 2,pattern color=.} \gpfillpath;
\gpcolor{rgb color={1.000,0.000,0.000}}
\draw[gp path] (5.207,3.907)--(6.123,3.907)--(6.123,4.061)--(5.207,4.061)--cycle;
\def\gpfillpath{(2.038,0.985)--(2.245,0.985)--(2.245,1.051)--(2.038,1.051)--cycle}
\gpfill{color=gpbgfillcolor} \gpfillpath;
\gpfill{rgb color={1.000,0.000,0.000},gp pattern 2,pattern color=.} \gpfillpath;
\draw[gp path] (2.038,0.985)--(2.038,1.050)--(2.244,1.050)--(2.244,0.985)--cycle;
\def\gpfillpath{(2.859,0.985)--(3.066,0.985)--(3.066,1.116)--(2.859,1.116)--cycle}
\gpfill{color=gpbgfillcolor} \gpfillpath;
\gpfill{rgb color={1.000,0.000,0.000},gp pattern 2,pattern color=.} \gpfillpath;
\draw[gp path] (2.859,0.985)--(2.859,1.115)--(3.065,1.115)--(3.065,0.985)--cycle;
\def\gpfillpath{(3.680,0.985)--(3.887,0.985)--(3.887,1.246)--(3.680,1.246)--cycle}
\gpfill{color=gpbgfillcolor} \gpfillpath;
\gpfill{rgb color={1.000,0.000,0.000},gp pattern 2,pattern color=.} \gpfillpath;
\draw[gp path] (3.680,0.985)--(3.680,1.245)--(3.886,1.245)--(3.886,0.985)--cycle;
\def\gpfillpath{(4.501,0.985)--(4.708,0.985)--(4.708,1.507)--(4.501,1.507)--cycle}
\gpfill{color=gpbgfillcolor} \gpfillpath;
\gpfill{rgb color={1.000,0.000,0.000},gp pattern 2,pattern color=.} \gpfillpath;
\draw[gp path] (4.501,0.985)--(4.501,1.506)--(4.707,1.506)--(4.707,0.985)--cycle;
\def\gpfillpath{(5.322,0.985)--(5.529,0.985)--(5.529,2.027)--(5.322,2.027)--cycle}
\gpfill{color=gpbgfillcolor} \gpfillpath;
\gpfill{rgb color={1.000,0.000,0.000},gp pattern 2,pattern color=.} \gpfillpath;
\draw[gp path] (5.322,0.985)--(5.322,2.026)--(5.528,2.026)--(5.528,0.985)--cycle;
\def\gpfillpath{(6.143,0.985)--(6.350,0.985)--(6.350,3.069)--(6.143,3.069)--cycle}
\gpfill{color=gpbgfillcolor} \gpfillpath;
\gpfill{rgb color={1.000,0.000,0.000},gp pattern 2,pattern color=.} \gpfillpath;
\draw[gp path] (6.143,0.985)--(6.143,3.068)--(6.349,3.068)--(6.349,0.985)--cycle;
\gpcolor{color=gp lt color border}
\node[gp node right] at (5.023,3.676) {Uni-source BFS};
\def\gpfillpath{(5.207,3.599)--(6.123,3.599)--(6.123,3.753)--(5.207,3.753)--cycle}
\gpfill{color=gpbgfillcolor} \gpfillpath;
\gpfill{rgb color={0.000,0.000,1.000},gp pattern 3,pattern color=.} \gpfillpath;
\gpcolor{rgb color={0.000,0.000,1.000}}
\draw[gp path] (5.207,3.599)--(6.123,3.599)--(6.123,3.753)--(5.207,3.753)--cycle;
\def\gpfillpath{(2.244,0.985)--(2.450,0.985)--(2.450,1.051)--(2.244,1.051)--cycle}
\gpfill{color=gpbgfillcolor} \gpfillpath;
\gpfill{rgb color={0.000,0.000,1.000},gp pattern 3,pattern color=.} \gpfillpath;
\draw[gp path] (2.244,0.985)--(2.244,1.050)--(2.449,1.050)--(2.449,0.985)--cycle;
\def\gpfillpath{(3.065,0.985)--(3.271,0.985)--(3.271,1.232)--(3.065,1.232)--cycle}
\gpfill{color=gpbgfillcolor} \gpfillpath;
\gpfill{rgb color={0.000,0.000,1.000},gp pattern 3,pattern color=.} \gpfillpath;
\draw[gp path] (3.065,0.985)--(3.065,1.231)--(3.270,1.231)--(3.270,0.985)--cycle;
\def\gpfillpath{(3.886,0.985)--(4.092,0.985)--(4.092,1.346)--(3.886,1.346)--cycle}
\gpfill{color=gpbgfillcolor} \gpfillpath;
\gpfill{rgb color={0.000,0.000,1.000},gp pattern 3,pattern color=.} \gpfillpath;
\draw[gp path] (3.886,0.985)--(3.886,1.345)--(4.091,1.345)--(4.091,0.985)--cycle;
\def\gpfillpath{(4.707,0.985)--(4.913,0.985)--(4.913,1.552)--(4.707,1.552)--cycle}
\gpfill{color=gpbgfillcolor} \gpfillpath;
\gpfill{rgb color={0.000,0.000,1.000},gp pattern 3,pattern color=.} \gpfillpath;
\draw[gp path] (4.707,0.985)--(4.707,1.551)--(4.912,1.551)--(4.912,0.985)--cycle;
\def\gpfillpath{(5.528,0.985)--(5.734,0.985)--(5.734,1.722)--(5.528,1.722)--cycle}
\gpfill{color=gpbgfillcolor} \gpfillpath;
\gpfill{rgb color={0.000,0.000,1.000},gp pattern 3,pattern color=.} \gpfillpath;
\draw[gp path] (5.528,0.985)--(5.528,1.721)--(5.733,1.721)--(5.733,0.985)--cycle;
\def\gpfillpath{(6.349,0.985)--(6.555,0.985)--(6.555,2.072)--(6.349,2.072)--cycle}
\gpfill{color=gpbgfillcolor} \gpfillpath;
\gpfill{rgb color={0.000,0.000,1.000},gp pattern 3,pattern color=.} \gpfillpath;
\draw[gp path] (6.349,0.985)--(6.349,2.071)--(6.554,2.071)--(6.554,0.985)--cycle;
\gpcolor{color=gp lt color border}
\draw[gp path] (1.320,3.393)--(1.320,0.985)--(7.067,0.985)--(7.067,3.393)--cycle;
\gpdefrectangularnode{gp plot 1}{\pgfpoint{1.320cm}{0.985cm}}{\pgfpoint{7.067cm}{3.393cm}}
\end{tikzpicture}

%% file: bc.tex
\subsection{Betweenness Centrality (BC)} \label{subsec:gt:bc}

Betweenness centrality measures the importance of a vertex in a network by
computing the number of shortest paths in which a vertex participates.
The most efficient algorithm to compute betweenness centrality \cite{bc}
is an iterative algorithm with computation complexity $O(nm + n^2 \log n)$,
for weighted graphs.

Betweenness centrality has three phases per iteration,
(i) breadth-first search (BFS) from a source vertex
(ii) backward propagation (BP), and
(iii) an accumulation phase (ACC). We derive the following principles:

\begin{figure}[ht]
    \centering
    \footnotesize
    \begin{subfigure}{\columnwidth}
        \centering
        \input{./charts/bc-cache}
        \caption{Multi-source and multi-source + async increase the
        ratio of cache hits per accessed page.}
        \label{fig:bc-cache}
    \end{subfigure}

    \begin{subfigure}{\columnwidth}
        \centering
        \input{./charts/bc-rt}
        \caption{Multi-source BC and multi-source + asnyc BC outperform multiple
        uni-source BC runs.}
        \label{fig:bc-rt}
    \end{subfigure}

    \caption[Multi-source betweenness centrality vs. uni-source.]{
        Multi-source asynchronous betweenness centrality
        compared with multiple uni-source and multi-source (synchronous) BC.}
    \label{fig:bc}
    \vspace{-10pt}
\end{figure}
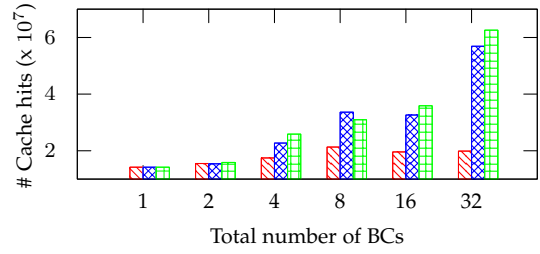
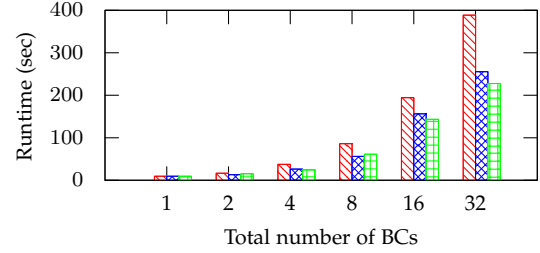

\textbf{Develop asynchronous applications}:
\graphyti{} adopts a multi-source betweenness centrality strategy,
similar to that of the graph diameter. The existence of 3 phases, provides the
opportunity for further optimization.
Eliminating phase synchrony for the multiple sources improves parallel efficiency.
Vertex activation messages now contain
metadata for both the current path(s) and the current phase(s).

\graphyti{}'s \texttt{betweenness} application separates algorithmic
design from the innate BSP paradigm within all vertex-centric frameworks.
Asynchronous design improves runtime by over $10\%$ when compared with just
multi-source and $40\%$ when compared to uni-source at $32$ sources
(Figure \ref{fig:bc}). Multi-source asynchronous betweenness
centrality reduces the amount of data
brought from disk by a factor of $4$ when $32$ concurrent searches are performed.

\textbf{Utilize functional constructs}:
Vertex-centric frameworks provide abstractions over threads that are accessible
to developers. Each partition thread in FlashGraph is a mechanism to
represent contention-free structures. As such, associative operations such
as functional reductions (e.g., \texttt{max}, \texttt{min}, \texttt{sum} etc.)
are naturally supported without resource contention. The BFS phase, computes a
global per-source-vertex \texttt{max}. The ACC phase, computes a global
per-source-vertex \texttt{add}. Both phases utilize this optimization.

%% file: charts/bc-cache.tex
\begin{tikzpicture}[gnuplot]
\path (0.000,0.000) rectangle (7.620,3.556);
\gpcolor{color=gp lt color border}
\gpsetlinetype{gp lt border}
\gpsetlinewidth{1.00}
\draw[gp path] (0.952,1.362)--(1.132,1.362);
\draw[gp path] (7.067,1.362)--(6.887,1.362);
\node[gp node right] at (0.768,1.362) {$2$};
\draw[gp path] (0.952,2.116)--(1.132,2.116);
\draw[gp path] (7.067,2.116)--(6.887,2.116);
\node[gp node right] at (0.768,2.116) {$4$};
\draw[gp path] (0.952,2.869)--(1.132,2.869);
\draw[gp path] (7.067,2.869)--(6.887,2.869);
\node[gp node right] at (0.768,2.869) {$6$};
\draw[gp path] (1.826,0.985)--(1.826,1.165);
\draw[gp path] (1.826,3.246)--(1.826,3.066);
\node[gp node center] at (1.826,0.677) {1};
\draw[gp path] (2.699,0.985)--(2.699,1.165);
\draw[gp path] (2.699,3.246)--(2.699,3.066);
\node[gp node center] at (2.699,0.677) {2};
\draw[gp path] (3.573,0.985)--(3.573,1.165);
\draw[gp path] (3.573,3.246)--(3.573,3.066);
\node[gp node center] at (3.573,0.677) {4};
\draw[gp path] (4.446,0.985)--(4.446,1.165);
\draw[gp path] (4.446,3.246)--(4.446,3.066);
\node[gp node center] at (4.446,0.677) {8};
\draw[gp path] (5.320,0.985)--(5.320,1.165);
\draw[gp path] (5.320,3.246)--(5.320,3.066);
\node[gp node center] at (5.320,0.677) {16};
\draw[gp path] (6.193,0.985)--(6.193,1.165);
\draw[gp path] (6.193,3.246)--(6.193,3.066);
\node[gp node center] at (6.193,0.677) {32};
\draw[gp path] (0.952,3.246)--(0.952,0.985)--(7.067,0.985)--(7.067,3.246)--cycle;
\node[gp node center,rotate=-270] at (0.276,2.115) {\# Cache hits (x $10^7$)};
\node[gp node center] at (4.009,0.215) {Total number of BCs};
\def\gpfillpath{(1.651,0.985)--(1.827,0.985)--(1.827,1.144)--(1.651,1.144)--cycle}
\gpfill{color=gpbgfillcolor} \gpfillpath;
\gpfill{rgb color={1.000,0.000,0.000},gp pattern 2,pattern color=.} \gpfillpath;
\gpcolor{rgb color={1.000,0.000,0.000}}
\draw[gp path] (1.651,0.985)--(1.651,1.143)--(1.826,1.143)--(1.826,0.985)--cycle;
\def\gpfillpath{(2.524,0.985)--(2.700,0.985)--(2.700,1.193)--(2.524,1.193)--cycle}
\gpfill{color=gpbgfillcolor} \gpfillpath;
\gpfill{rgb color={1.000,0.000,0.000},gp pattern 2,pattern color=.} \gpfillpath;
\draw[gp path] (2.524,0.985)--(2.524,1.192)--(2.699,1.192)--(2.699,0.985)--cycle;
\def\gpfillpath{(3.398,0.985)--(3.574,0.985)--(3.574,1.267)--(3.398,1.267)--cycle}
\gpfill{color=gpbgfillcolor} \gpfillpath;
\gpfill{rgb color={1.000,0.000,0.000},gp pattern 2,pattern color=.} \gpfillpath;
\draw[gp path] (3.398,0.985)--(3.398,1.266)--(3.573,1.266)--(3.573,0.985)--cycle;
\def\gpfillpath{(4.272,0.985)--(4.447,0.985)--(4.447,1.412)--(4.272,1.412)--cycle}
\gpfill{color=gpbgfillcolor} \gpfillpath;
\gpfill{rgb color={1.000,0.000,0.000},gp pattern 2,pattern color=.} \gpfillpath;
\draw[gp path] (4.272,0.985)--(4.272,1.411)--(4.446,1.411)--(4.446,0.985)--cycle;
\def\gpfillpath{(5.145,0.985)--(5.321,0.985)--(5.321,1.347)--(5.145,1.347)--cycle}
\gpfill{color=gpbgfillcolor} \gpfillpath;
\gpfill{rgb color={1.000,0.000,0.000},gp pattern 2,pattern color=.} \gpfillpath;
\draw[gp path] (5.145,0.985)--(5.145,1.346)--(5.320,1.346)--(5.320,0.985)--cycle;
\def\gpfillpath{(6.019,0.985)--(6.194,0.985)--(6.194,1.358)--(6.019,1.358)--cycle}
\gpfill{color=gpbgfillcolor} \gpfillpath;
\gpfill{rgb color={1.000,0.000,0.000},gp pattern 2,pattern color=.} \gpfillpath;
\draw[gp path] (6.019,0.985)--(6.019,1.357)--(6.193,1.357)--(6.193,0.985)--cycle;
\def\gpfillpath{(1.826,0.985)--(2.001,0.985)--(2.001,1.144)--(1.826,1.144)--cycle}
\gpfill{color=gpbgfillcolor} \gpfillpath;
\gpfill{rgb color={0.000,0.000,1.000},gp pattern 3,pattern color=.} \gpfillpath;
\gpcolor{rgb color={0.000,0.000,1.000}}
\draw[gp path] (1.826,0.985)--(1.826,1.143)--(2.000,1.143)--(2.000,0.985)--cycle;
\def\gpfillpath{(2.699,0.985)--(2.875,0.985)--(2.875,1.189)--(2.699,1.189)--cycle}
\gpfill{color=gpbgfillcolor} \gpfillpath;
\gpfill{rgb color={0.000,0.000,1.000},gp pattern 3,pattern color=.} \gpfillpath;
\draw[gp path] (2.699,0.985)--(2.699,1.188)--(2.874,1.188)--(2.874,0.985)--cycle;
\def\gpfillpath{(3.573,0.985)--(3.748,0.985)--(3.748,1.465)--(3.573,1.465)--cycle}
\gpfill{color=gpbgfillcolor} \gpfillpath;
\gpfill{rgb color={0.000,0.000,1.000},gp pattern 3,pattern color=.} \gpfillpath;
\draw[gp path] (3.573,0.985)--(3.573,1.464)--(3.747,1.464)--(3.747,0.985)--cycle;
\def\gpfillpath{(4.446,0.985)--(4.622,0.985)--(4.622,1.876)--(4.446,1.876)--cycle}
\gpfill{color=gpbgfillcolor} \gpfillpath;
\gpfill{rgb color={0.000,0.000,1.000},gp pattern 3,pattern color=.} \gpfillpath;
\draw[gp path] (4.446,0.985)--(4.446,1.875)--(4.621,1.875)--(4.621,0.985)--cycle;
\def\gpfillpath{(5.320,0.985)--(5.496,0.985)--(5.496,1.838)--(5.320,1.838)--cycle}
\gpfill{color=gpbgfillcolor} \gpfillpath;
\gpfill{rgb color={0.000,0.000,1.000},gp pattern 3,pattern color=.} \gpfillpath;
\draw[gp path] (5.320,0.985)--(5.320,1.837)--(5.495,1.837)--(5.495,0.985)--cycle;
\def\gpfillpath{(6.193,0.985)--(6.369,0.985)--(6.369,2.753)--(6.193,2.753)--cycle}
\gpfill{color=gpbgfillcolor} \gpfillpath;
\gpfill{rgb color={0.000,0.000,1.000},gp pattern 3,pattern color=.} \gpfillpath;
\draw[gp path] (6.193,0.985)--(6.193,2.752)--(6.368,2.752)--(6.368,0.985)--cycle;
\def\gpfillpath{(2.000,0.985)--(2.176,0.985)--(2.176,1.144)--(2.000,1.144)--cycle}
\gpfill{color=gpbgfillcolor} \gpfillpath;
\gpfill{rgb color={0.000,1.000,0.000},gp pattern 4,pattern color=.} \gpfillpath;
\gpcolor{rgb color={0.000,1.000,0.000}}
\draw[gp path] (2.000,0.985)--(2.000,1.143)--(2.175,1.143)--(2.175,0.985)--cycle;
\def\gpfillpath{(2.874,0.985)--(3.050,0.985)--(3.050,1.207)--(2.874,1.207)--cycle}
\gpfill{color=gpbgfillcolor} \gpfillpath;
\gpfill{rgb color={0.000,1.000,0.000},gp pattern 4,pattern color=.} \gpfillpath;
\draw[gp path] (2.874,0.985)--(2.874,1.206)--(3.049,1.206)--(3.049,0.985)--cycle;
\def\gpfillpath{(3.747,0.985)--(3.923,0.985)--(3.923,1.584)--(3.747,1.584)--cycle}
\gpfill{color=gpbgfillcolor} \gpfillpath;
\gpfill{rgb color={0.000,1.000,0.000},gp pattern 4,pattern color=.} \gpfillpath;
\draw[gp path] (3.747,0.985)--(3.747,1.583)--(3.922,1.583)--(3.922,0.985)--cycle;
\def\gpfillpath{(4.621,0.985)--(4.797,0.985)--(4.797,1.775)--(4.621,1.775)--cycle}
\gpfill{color=gpbgfillcolor} \gpfillpath;
\gpfill{rgb color={0.000,1.000,0.000},gp pattern 4,pattern color=.} \gpfillpath;
\draw[gp path] (4.621,0.985)--(4.621,1.774)--(4.796,1.774)--(4.796,0.985)--cycle;
\def\gpfillpath{(5.495,0.985)--(5.670,0.985)--(5.670,1.961)--(5.495,1.961)--cycle}
\gpfill{color=gpbgfillcolor} \gpfillpath;
\gpfill{rgb color={0.000,1.000,0.000},gp pattern 4,pattern color=.} \gpfillpath;
\draw[gp path] (5.495,0.985)--(5.495,1.960)--(5.669,1.960)--(5.669,0.985)--cycle;
\def\gpfillpath{(6.368,0.985)--(6.544,0.985)--(6.544,2.967)--(6.368,2.967)--cycle}
\gpfill{color=gpbgfillcolor} \gpfillpath;
\gpfill{rgb color={0.000,1.000,0.000},gp pattern 4,pattern color=.} \gpfillpath;
\draw[gp path] (6.368,0.985)--(6.368,2.966)--(6.543,2.966)--(6.543,0.985)--cycle;
\gpcolor{color=gp lt color border}
\draw[gp path] (0.952,3.246)--(0.952,0.985)--(7.067,0.985)--(7.067,3.246)--cycle;
\gpdefrectangularnode{gp plot 1}{\pgfpoint{0.952cm}{0.985cm}}{\pgfpoint{7.067cm}{3.246cm}}
\end{tikzpicture}

%% file: charts/bc-rt.tex
\begin{tikzpicture}[gnuplot]
\path (0.000,0.000) rectangle (7.620,3.556);
\gpcolor{color=gp lt color border}
\gpsetlinetype{gp lt border}
\gpsetlinewidth{1.00}
\draw[gp path] (1.320,0.985)--(1.500,0.985);
\draw[gp path] (7.067,0.985)--(6.887,0.985);
\node[gp node right] at (1.136,0.985) {$0$};
\draw[gp path] (1.320,1.550)--(1.500,1.550);
\draw[gp path] (7.067,1.550)--(6.887,1.550);
\node[gp node right] at (1.136,1.550) {$100$};
\draw[gp path] (1.320,2.116)--(1.500,2.116);
\draw[gp path] (7.067,2.116)--(6.887,2.116);
\node[gp node right] at (1.136,2.116) {$200$};
\draw[gp path] (1.320,2.681)--(1.500,2.681);
\draw[gp path] (7.067,2.681)--(6.887,2.681);
\node[gp node right] at (1.136,2.681) {$300$};
\draw[gp path] (1.320,3.246)--(1.500,3.246);
\draw[gp path] (7.067,3.246)--(6.887,3.246);
\node[gp node right] at (1.136,3.246) {$400$};
\draw[gp path] (2.141,0.985)--(2.141,1.165);
\draw[gp path] (2.141,3.246)--(2.141,3.066);
\node[gp node center] at (2.141,0.677) {1};
\draw[gp path] (2.962,0.985)--(2.962,1.165);
\draw[gp path] (2.962,3.246)--(2.962,3.066);
\node[gp node center] at (2.962,0.677) {2};
\draw[gp path] (3.783,0.985)--(3.783,1.165);
\draw[gp path] (3.783,3.246)--(3.783,3.066);
\node[gp node center] at (3.783,0.677) {4};
\draw[gp path] (4.604,0.985)--(4.604,1.165);
\draw[gp path] (4.604,3.246)--(4.604,3.066);
\node[gp node center] at (4.604,0.677) {8};
\draw[gp path] (5.425,0.985)--(5.425,1.165);
\draw[gp path] (5.425,3.246)--(5.425,3.066);
\node[gp node center] at (5.425,0.677) {16};
\draw[gp path] (6.246,0.985)--(6.246,1.165);
\draw[gp path] (6.246,3.246)--(6.246,3.066);
\node[gp node center] at (6.246,0.677) {32};
\draw[gp path] (1.320,3.246)--(1.320,0.985)--(7.067,0.985)--(7.067,3.246)--cycle;
\node[gp node center,rotate=-270] at (0.276,2.115) {Runtime (sec)};
\node[gp node center] at (4.193,0.215) {Total number of BCs};
\def\gpfillpath{(1.977,0.985)--(2.142,0.985)--(2.142,1.039)--(1.977,1.039)--cycle}
\gpfill{color=gpbgfillcolor} \gpfillpath;
\gpfill{rgb color={1.000,0.000,0.000},gp pattern 2,pattern color=.} \gpfillpath;
\gpcolor{rgb color={1.000,0.000,0.000}}
\draw[gp path] (1.977,0.985)--(1.977,1.038)--(2.141,1.038)--(2.141,0.985)--cycle;
\def\gpfillpath{(2.798,0.985)--(2.963,0.985)--(2.963,1.079)--(2.798,1.079)--cycle}
\gpfill{color=gpbgfillcolor} \gpfillpath;
\gpfill{rgb color={1.000,0.000,0.000},gp pattern 2,pattern color=.} \gpfillpath;
\draw[gp path] (2.798,0.985)--(2.798,1.078)--(2.962,1.078)--(2.962,0.985)--cycle;
\def\gpfillpath{(3.619,0.985)--(3.784,0.985)--(3.784,1.197)--(3.619,1.197)--cycle}
\gpfill{color=gpbgfillcolor} \gpfillpath;
\gpfill{rgb color={1.000,0.000,0.000},gp pattern 2,pattern color=.} \gpfillpath;
\draw[gp path] (3.619,0.985)--(3.619,1.196)--(3.783,1.196)--(3.783,0.985)--cycle;
\def\gpfillpath{(4.440,0.985)--(4.605,0.985)--(4.605,1.472)--(4.440,1.472)--cycle}
\gpfill{color=gpbgfillcolor} \gpfillpath;
\gpfill{rgb color={1.000,0.000,0.000},gp pattern 2,pattern color=.} \gpfillpath;
\draw[gp path] (4.440,0.985)--(4.440,1.471)--(4.604,1.471)--(4.604,0.985)--cycle;
\def\gpfillpath{(5.261,0.985)--(5.426,0.985)--(5.426,2.084)--(5.261,2.084)--cycle}
\gpfill{color=gpbgfillcolor} \gpfillpath;
\gpfill{rgb color={1.000,0.000,0.000},gp pattern 2,pattern color=.} \gpfillpath;
\draw[gp path] (5.261,0.985)--(5.261,2.083)--(5.425,2.083)--(5.425,0.985)--cycle;
\def\gpfillpath{(6.082,0.985)--(6.247,0.985)--(6.247,3.182)--(6.082,3.182)--cycle}
\gpfill{color=gpbgfillcolor} \gpfillpath;
\gpfill{rgb color={1.000,0.000,0.000},gp pattern 2,pattern color=.} \gpfillpath;
\draw[gp path] (6.082,0.985)--(6.082,3.181)--(6.246,3.181)--(6.246,0.985)--cycle;
\def\gpfillpath{(2.141,0.985)--(2.306,0.985)--(2.306,1.039)--(2.141,1.039)--cycle}
\gpfill{color=gpbgfillcolor} \gpfillpath;
\gpfill{rgb color={0.000,0.000,1.000},gp pattern 3,pattern color=.} \gpfillpath;
\gpcolor{rgb color={0.000,0.000,1.000}}
\draw[gp path] (2.141,0.985)--(2.141,1.038)--(2.305,1.038)--(2.305,0.985)--cycle;
\def\gpfillpath{(2.962,0.985)--(3.127,0.985)--(3.127,1.060)--(2.962,1.060)--cycle}
\gpfill{color=gpbgfillcolor} \gpfillpath;
\gpfill{rgb color={0.000,0.000,1.000},gp pattern 3,pattern color=.} \gpfillpath;
\draw[gp path] (2.962,0.985)--(2.962,1.059)--(3.126,1.059)--(3.126,0.985)--cycle;
\def\gpfillpath{(3.783,0.985)--(3.948,0.985)--(3.948,1.134)--(3.783,1.134)--cycle}
\gpfill{color=gpbgfillcolor} \gpfillpath;
\gpfill{rgb color={0.000,0.000,1.000},gp pattern 3,pattern color=.} \gpfillpath;
\draw[gp path] (3.783,0.985)--(3.783,1.133)--(3.947,1.133)--(3.947,0.985)--cycle;
\def\gpfillpath{(4.604,0.985)--(4.769,0.985)--(4.769,1.303)--(4.604,1.303)--cycle}
\gpfill{color=gpbgfillcolor} \gpfillpath;
\gpfill{rgb color={0.000,0.000,1.000},gp pattern 3,pattern color=.} \gpfillpath;
\draw[gp path] (4.604,0.985)--(4.604,1.302)--(4.768,1.302)--(4.768,0.985)--cycle;
\def\gpfillpath{(5.425,0.985)--(5.590,0.985)--(5.590,1.869)--(5.425,1.869)--cycle}
\gpfill{color=gpbgfillcolor} \gpfillpath;
\gpfill{rgb color={0.000,0.000,1.000},gp pattern 3,pattern color=.} \gpfillpath;
\draw[gp path] (5.425,0.985)--(5.425,1.868)--(5.589,1.868)--(5.589,0.985)--cycle;
\def\gpfillpath{(6.246,0.985)--(6.411,0.985)--(6.411,2.430)--(6.246,2.430)--cycle}
\gpfill{color=gpbgfillcolor} \gpfillpath;
\gpfill{rgb color={0.000,0.000,1.000},gp pattern 3,pattern color=.} \gpfillpath;
\draw[gp path] (6.246,0.985)--(6.246,2.429)--(6.410,2.429)--(6.410,0.985)--cycle;
\def\gpfillpath{(2.305,0.985)--(2.470,0.985)--(2.470,1.039)--(2.305,1.039)--cycle}
\gpfill{color=gpbgfillcolor} \gpfillpath;
\gpfill{rgb color={0.000,1.000,0.000},gp pattern 4,pattern color=.} \gpfillpath;
\gpcolor{rgb color={0.000,1.000,0.000}}
\draw[gp path] (2.305,0.985)--(2.305,1.038)--(2.469,1.038)--(2.469,0.985)--cycle;
\def\gpfillpath{(3.126,0.985)--(3.291,0.985)--(3.291,1.072)--(3.126,1.072)--cycle}
\gpfill{color=gpbgfillcolor} \gpfillpath;
\gpfill{rgb color={0.000,1.000,0.000},gp pattern 4,pattern color=.} \gpfillpath;
\draw[gp path] (3.126,0.985)--(3.126,1.071)--(3.290,1.071)--(3.290,0.985)--cycle;
\def\gpfillpath{(3.947,0.985)--(4.112,0.985)--(4.112,1.123)--(3.947,1.123)--cycle}
\gpfill{color=gpbgfillcolor} \gpfillpath;
\gpfill{rgb color={0.000,1.000,0.000},gp pattern 4,pattern color=.} \gpfillpath;
\draw[gp path] (3.947,0.985)--(3.947,1.122)--(4.111,1.122)--(4.111,0.985)--cycle;
\def\gpfillpath{(4.768,0.985)--(4.933,0.985)--(4.933,1.332)--(4.768,1.332)--cycle}
\gpfill{color=gpbgfillcolor} \gpfillpath;
\gpfill{rgb color={0.000,1.000,0.000},gp pattern 4,pattern color=.} \gpfillpath;
\draw[gp path] (4.768,0.985)--(4.768,1.331)--(4.932,1.331)--(4.932,0.985)--cycle;
\def\gpfillpath{(5.589,0.985)--(5.754,0.985)--(5.754,1.796)--(5.589,1.796)--cycle}
\gpfill{color=gpbgfillcolor} \gpfillpath;
\gpfill{rgb color={0.000,1.000,0.000},gp pattern 4,pattern color=.} \gpfillpath;
\draw[gp path] (5.589,0.985)--(5.589,1.795)--(5.753,1.795)--(5.753,0.985)--cycle;
\def\gpfillpath{(6.410,0.985)--(6.575,0.985)--(6.575,2.273)--(6.410,2.273)--cycle}
\gpfill{color=gpbgfillcolor} \gpfillpath;
\gpfill{rgb color={0.000,1.000,0.000},gp pattern 4,pattern color=.} \gpfillpath;
\draw[gp path] (6.410,0.985)--(6.410,2.272)--(6.574,2.272)--(6.574,0.985)--cycle;
\gpcolor{color=gp lt color border}
\draw[gp path] (1.320,3.246)--(1.320,0.985)--(7.067,0.985)--(7.067,3.246)--cycle;
\gpdefrectangularnode{gp plot 1}{\pgfpoint{1.320cm}{0.985cm}}{\pgfpoint{7.067cm}{3.246cm}}
\end{tikzpicture}

%% file: tc.tex
\subsection{Triangle Counting} \label{subsec:gt:tc}

Triangle counting is a topological structure discovery algorithm that
finds pairs of vertices that share a common neighbor.
When performed in SEM, the complexity is $\mathcal{O}(n^3)$.
The fundamental task finds the intersection of the adjacency lists of
neighboring vertices to discover triangles. 
Each vertex requests its neighbor's adjacency lists and computes when the list
is available in the page cache.
We discount alternative
implementations in which the state of a vertex can exceed the size of its own
edge list and that of one other neighbor because they would violate the SEM
limited memory usage guarantee.

\textbf{Optimize in-memory operations}:
Once data has been brought into memory it is essential to not only reuse cached
data, but perform in-memory optimizations.
The following accelerates the intersection search operation:
\begin{itemize}
    \item Store adjacency lists in sorted order. This allows the implementation 
        to choose between binary search and sequential scans when appropriate.
    \item Store the adjacency list of a vertex with degree higher than a certain
        threshold in a hash table to improve lookup performance.
    \item Perform a \textit{restarted} binary search in the event an element is
        not found. A restarted binary search looks for the next item using the
        end point of the previous search. 
    \item Order the adjacency list enumeration appropriately.
        This choice will lead to either forward or reverse traversal
        of edge lists being more efficient. In our case, reverse
        iteration leads to an improvement of $1.7 X$ in search. This is because
        the discovery of triangles is performed by higher degree vertices
        leading to fewer requests for edge lists of lower degree vertices.
\end{itemize}

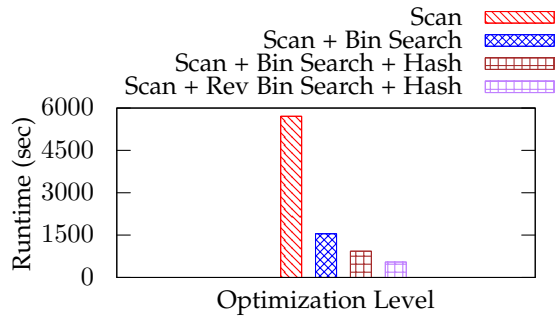
\begin{figure}[!hbt]
    \centering
        \input{./charts/tc}
    \caption[Triangle counting incremental optimizations.]{
        Incremental optimizations applied to triangle counting.
        All optimizations improve performance by two orders of magnitude 
        when compared with a scan adjacency list intersection.}
    \label{fig:tc}
\end{figure}

Figure \ref{fig:tc} displays the improvement from each of the in-memory
optimizations. After all optimizations are applied, \graphyti{}'s triangle
counting performs two orders of magnitude faster on average.

%% file: charts/tc.tex
\begin{tikzpicture}[gnuplot]
\path (0.000,0.000) rectangle (7.620,4.318);
\gpcolor{color=gp lt color border}
\gpsetlinetype{gp lt border}
\gpsetlinewidth{1.00}
\draw[gp path] (1.504,0.523)--(1.684,0.523);
\draw[gp path] (7.067,0.523)--(6.887,0.523);
\node[gp node right] at (1.320,0.523) {$0$};
\draw[gp path] (1.504,1.087)--(1.684,1.087);
\draw[gp path] (7.067,1.087)--(6.887,1.087);
\node[gp node right] at (1.320,1.087) {$1500$};
\draw[gp path] (1.504,1.650)--(1.684,1.650);
\draw[gp path] (7.067,1.650)--(6.887,1.650);
\node[gp node right] at (1.320,1.650) {$3000$};
\draw[gp path] (1.504,2.214)--(1.684,2.214);
\draw[gp path] (7.067,2.214)--(6.887,2.214);
\node[gp node right] at (1.320,2.214) {$4500$};
\draw[gp path] (1.504,2.777)--(1.684,2.777);
\draw[gp path] (7.067,2.777)--(6.887,2.777);
\node[gp node right] at (1.320,2.777) {$6000$};
\draw[gp path] (1.504,2.777)--(1.504,0.523)--(7.067,0.523)--(7.067,2.777)--cycle;
\node[gp node center,rotate=-270] at (0.276,1.650) {Runtime (sec)};
\node[gp node center] at (4.285,0.215) {Optimization Level};
\node[gp node right] at (6.219,3.984) {Scan};
\def\gpfillpath{(6.403,3.907)--(7.319,3.907)--(7.319,4.061)--(6.403,4.061)--cycle}
\gpfill{color=gpbgfillcolor} \gpfillpath;
\gpfill{rgb color={1.000,0.000,0.000},gp pattern 2,pattern color=.} \gpfillpath;
\gpcolor{rgb color={1.000,0.000,0.000}}
\draw[gp path] (6.403,3.907)--(7.319,3.907)--(7.319,4.061)--(6.403,4.061)--cycle;
\def\gpfillpath{(3.683,0.523)--(3.962,0.523)--(3.962,2.668)--(3.683,2.668)--cycle}
\gpfill{color=gpbgfillcolor} \gpfillpath;
\gpfill{rgb color={1.000,0.000,0.000},gp pattern 2,pattern color=.} \gpfillpath;
\draw[gp path] (3.683,0.523)--(3.683,2.667)--(3.961,2.667)--(3.961,0.523)--cycle;
\gpcolor{color=gp lt color border}
\node[gp node right] at (6.219,3.676) {Scan + Bin Search};
\def\gpfillpath{(6.403,3.599)--(7.319,3.599)--(7.319,3.753)--(6.403,3.753)--cycle}
\gpfill{color=gpbgfillcolor} \gpfillpath;
\gpfill{rgb color={0.000,0.000,1.000},gp pattern 3,pattern color=.} \gpfillpath;
\gpcolor{rgb color={0.000,0.000,1.000}}
\draw[gp path] (6.403,3.599)--(7.319,3.599)--(7.319,3.753)--(6.403,3.753)--cycle;
\def\gpfillpath{(4.146,0.523)--(4.426,0.523)--(4.426,1.105)--(4.146,1.105)--cycle}
\gpfill{color=gpbgfillcolor} \gpfillpath;
\gpfill{rgb color={0.000,0.000,1.000},gp pattern 3,pattern color=.} \gpfillpath;
\draw[gp path] (4.146,0.523)--(4.146,1.104)--(4.425,1.104)--(4.425,0.523)--cycle;
\gpcolor{color=gp lt color border}
\node[gp node right] at (6.219,3.368) {Scan + Bin Search + Hash};
\def\gpfillpath{(6.403,3.291)--(7.319,3.291)--(7.319,3.445)--(6.403,3.445)--cycle}
\gpfill{color=gpbgfillcolor} \gpfillpath;
\gpfill{rgb color={0.647,0.165,0.165},gp pattern 4,pattern color=.} \gpfillpath;
\gpcolor{rgb color={0.647,0.165,0.165}}
\draw[gp path] (6.403,3.291)--(7.319,3.291)--(7.319,3.445)--(6.403,3.445)--cycle;
\def\gpfillpath{(4.610,0.523)--(4.889,0.523)--(4.889,0.873)--(4.610,0.873)--cycle}
\gpfill{color=gpbgfillcolor} \gpfillpath;
\gpfill{rgb color={0.647,0.165,0.165},gp pattern 4,pattern color=.} \gpfillpath;
\draw[gp path] (4.610,0.523)--(4.610,0.872)--(4.888,0.872)--(4.888,0.523)--cycle;
\gpcolor{color=gp lt color border}
\node[gp node right] at (6.219,3.060) {Scan + Rev Bin Search + Hash};
\def\gpfillpath{(6.403,2.983)--(7.319,2.983)--(7.319,3.137)--(6.403,3.137)--cycle}
\gpfill{color=gpbgfillcolor} \gpfillpath;
\gpfill{rgb color={0.753,0.502,1.000},gp pattern 4,pattern color=.} \gpfillpath;
\gpcolor{rgb color={0.753,0.502,1.000}}
\draw[gp path] (6.403,2.983)--(7.319,2.983)--(7.319,3.137)--(6.403,3.137)--cycle;
\def\gpfillpath{(5.074,0.523)--(5.353,0.523)--(5.353,0.729)--(5.074,0.729)--cycle}
\gpfill{color=gpbgfillcolor} \gpfillpath;
\gpfill{rgb color={0.753,0.502,1.000},gp pattern 4,pattern color=.} \gpfillpath;
\draw[gp path] (5.074,0.523)--(5.074,0.728)--(5.352,0.728)--(5.352,0.523)--cycle;
\gpcolor{color=gp lt color border}
\draw[gp path] (1.504,2.777)--(1.504,0.523)--(7.067,0.523)--(7.067,2.777)--cycle;
\gpdefrectangularnode{gp plot 1}{\pgfpoint{1.504cm}{0.523cm}}{\pgfpoint{7.067cm}{2.777cm}}
\end{tikzpicture}

%% file: louvain.tex
\subsection{Louvain Modularity} \label{subsec:gt:louvain}

Louvain modularity \cite{louvain} is an agglomerative community detection
algorithm that maximizes the density of edges within communities and minimizes those outside.
Modularity for any pair of communities $i$ and $j$ is computed as:
\begin{equation} \label{eqn:louvain-mod}
    Q = \frac{1}{2m} \sum_{ij} (A_{ij} - \frac{k_i k_j}{2m}) \delta(c_i, c_j),
\end{equation}

\noindent in which $m$ is the sum of all graph edge weights,
$A_{ij}$ is the edge weight between $v_i$ and $v_j$,
$k_i$ and $k_j$ are the weighted sum of edges between $v_i$ and $v_j$,
$\delta$ is a function that differentiates one community from the next.

We start with the most popular two phase, greedy approximation algorithm \cite{louvain}.
Exact solutions are computationally infeasible for large networks.
A vertex changes community to another that contains the maximum
\textbf{positive} modularity among neighboring communities.



\begin{figure}[!h]
    \centering
    \footnotesize
    \begin{subfigure}{\columnwidth}
        \centering
        \input{./charts/louvain}
        \caption[Optimized Louvain]{The breakdown of runtime for \graphyti{}'s
        \texttt{louvain}.}
        \label{fig:louvain-opt}
    \end{subfigure}

    \begin{subfigure}{\columnwidth}
        \centering
        \input{./charts/louvain-naive}
        \caption[Unoptimized Louvain]{Performance of Louvain
        when computed through progressive materializations of communities.}
        \label{fig:louvain-naive}
    \end{subfigure}

        \caption[Optimized Louvain vs. materialized, traditional
        Louvain]{\graphyti{}'s \texttt{louvain} runs $2$ X faster
        than the traditional implementation with physical graph modifications.}
    \label{fig:louvain}
\end{figure}
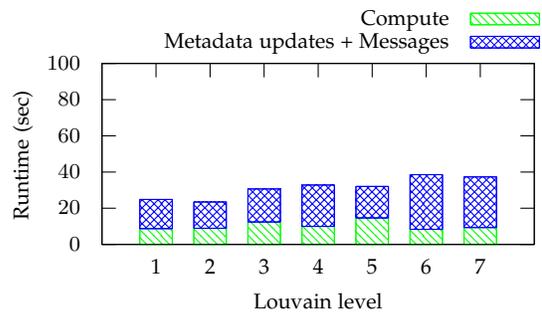
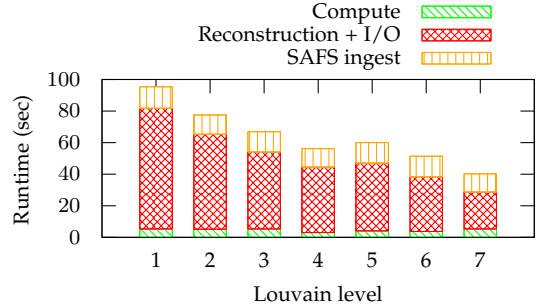

This algorithm poses challenges for SEM frameworks because it modifies the graph
structure. Modification is extremely expensive, because edge data need to be rewritten on disk.
We adopt the following principle:

\textbf{Avoid graph structure modification}:
Modifying the graph is prohibitively expensive. In
fact, for SEM applications, modification can easily surpass the algorithmic
runtime,  because disk write throughput is orders of magnitude slower
than memory throughput. We demonstrate this in Figure \ref{fig:louvain-naive}.
Accordingly, we circumvent modification through (i) lazy
deletion and (ii) vertex nomination of a \textit{community representative}.
We maintain a partitioned bitmap with lookups for
deleted vertices in addition to an index for vertex-to-community lookups. This
ensures all messages are appropriately routed to the correct vertex without
involving the graph engine or requiring messages to be forwarded.

Figure \ref{fig:louvain-naive} shows the ``best-case scenario'' for an SEM
implementation that physically modifies the graph. We maintain a RAMDisk in
fast DDR$4$ to hold the new physical state of the graph.
\graphyti{}'s \texttt{louvain} performs twice as fast as this best case
(Figure \ref{fig:louvain-opt}). We trade-off graph structure modification with
metadata updates and messaging. Naturally, as the algorithm progresses to
deeper levels, more vertices merge, resulting in fewer clusters.
This reduces the cost of traditional graph modification, while conversely
increasing the overhead of messaging and metadata maintenance for \graphyti{}'s
\texttt{louvain}. Accordingly, \graphyti{}'s \texttt{louvain} design capitalizes
most during early levels to attain its performance gains.

%% file: charts/louvain.tex
\begin{tikzpicture}[gnuplot]
\path (0.000,0.000) rectangle (7.620,4.318);
\gpcolor{color=gp lt color border}
\gpsetlinetype{gp lt border}
\gpsetlinewidth{1.00}
\draw[gp path] (1.320,0.985)--(1.500,0.985);
\draw[gp path] (7.067,0.985)--(6.887,0.985);
\node[gp node right] at (1.136,0.985) {$0$};
\draw[gp path] (1.320,1.467)--(1.500,1.467);
\draw[gp path] (7.067,1.467)--(6.887,1.467);
\node[gp node right] at (1.136,1.467) {$20$};
\draw[gp path] (1.320,1.948)--(1.500,1.948);
\draw[gp path] (7.067,1.948)--(6.887,1.948);
\node[gp node right] at (1.136,1.948) {$40$};
\draw[gp path] (1.320,2.430)--(1.500,2.430);
\draw[gp path] (7.067,2.430)--(6.887,2.430);
\node[gp node right] at (1.136,2.430) {$60$};
\draw[gp path] (1.320,2.911)--(1.500,2.911);
\draw[gp path] (7.067,2.911)--(6.887,2.911);
\node[gp node right] at (1.136,2.911) {$80$};
\draw[gp path] (1.320,3.393)--(1.500,3.393);
\draw[gp path] (7.067,3.393)--(6.887,3.393);
\node[gp node right] at (1.136,3.393) {$100$};
\draw[gp path] (2.038,0.985)--(2.038,1.165);
\draw[gp path] (2.038,3.393)--(2.038,3.213);
\node[gp node center] at (2.038,0.677) {1};
\draw[gp path] (2.757,0.985)--(2.757,1.165);
\draw[gp path] (2.757,3.393)--(2.757,3.213);
\node[gp node center] at (2.757,0.677) {2};
\draw[gp path] (3.475,0.985)--(3.475,1.165);
\draw[gp path] (3.475,3.393)--(3.475,3.213);
\node[gp node center] at (3.475,0.677) {3};
\draw[gp path] (4.194,0.985)--(4.194,1.165);
\draw[gp path] (4.194,3.393)--(4.194,3.213);
\node[gp node center] at (4.194,0.677) {4};
\draw[gp path] (4.912,0.985)--(4.912,1.165);
\draw[gp path] (4.912,3.393)--(4.912,3.213);
\node[gp node center] at (4.912,0.677) {5};
\draw[gp path] (5.630,0.985)--(5.630,1.165);
\draw[gp path] (5.630,3.393)--(5.630,3.213);
\node[gp node center] at (5.630,0.677) {6};
\draw[gp path] (6.349,0.985)--(6.349,1.165);
\draw[gp path] (6.349,3.393)--(6.349,3.213);
\node[gp node center] at (6.349,0.677) {7};
\draw[gp path] (1.320,3.393)--(1.320,0.985)--(7.067,0.985)--(7.067,3.393)--cycle;
\node[gp node center,rotate=-270] at (0.276,2.189) {Runtime (sec)};
\node[gp node center] at (4.193,0.215) {Louvain level};
\node[gp node right] at (6.035,3.984) {Compute};
\def\gpfillpath{(6.219,3.907)--(7.135,3.907)--(7.135,4.061)--(6.219,4.061)--cycle}
\gpfill{color=gpbgfillcolor} \gpfillpath;
\gpfill{rgb color={0.000,1.000,0.000},gp pattern 2,pattern color=.} \gpfillpath;
\gpcolor{rgb color={0.000,1.000,0.000}}
\draw[gp path] (6.219,3.907)--(7.135,3.907)--(7.135,4.061)--(6.219,4.061)--cycle;
\def\gpfillpath{(1.823,0.985)--(2.255,0.985)--(2.255,1.196)--(1.823,1.196)--cycle}
\gpfill{color=gpbgfillcolor} \gpfillpath;
\gpfill{rgb color={0.000,1.000,0.000},gp pattern 2,pattern color=.} \gpfillpath;
\draw[gp path] (1.823,0.985)--(1.823,1.195)--(2.254,1.195)--(2.254,0.985)--cycle;
\def\gpfillpath{(2.541,0.985)--(2.973,0.985)--(2.973,1.201)--(2.541,1.201)--cycle}
\gpfill{color=gpbgfillcolor} \gpfillpath;
\gpfill{rgb color={0.000,1.000,0.000},gp pattern 2,pattern color=.} \gpfillpath;
\draw[gp path] (2.541,0.985)--(2.541,1.200)--(2.972,1.200)--(2.972,0.985)--cycle;
\def\gpfillpath{(3.260,0.985)--(3.692,0.985)--(3.692,1.285)--(3.260,1.285)--cycle}
\gpfill{color=gpbgfillcolor} \gpfillpath;
\gpfill{rgb color={0.000,1.000,0.000},gp pattern 2,pattern color=.} \gpfillpath;
\draw[gp path] (3.260,0.985)--(3.260,1.284)--(3.691,1.284)--(3.691,0.985)--cycle;
\def\gpfillpath{(3.978,0.985)--(4.410,0.985)--(4.410,1.226)--(3.978,1.226)--cycle}
\gpfill{color=gpbgfillcolor} \gpfillpath;
\gpfill{rgb color={0.000,1.000,0.000},gp pattern 2,pattern color=.} \gpfillpath;
\draw[gp path] (3.978,0.985)--(3.978,1.225)--(4.409,1.225)--(4.409,0.985)--cycle;
\def\gpfillpath{(4.696,0.985)--(5.128,0.985)--(5.128,1.339)--(4.696,1.339)--cycle}
\gpfill{color=gpbgfillcolor} \gpfillpath;
\gpfill{rgb color={0.000,1.000,0.000},gp pattern 2,pattern color=.} \gpfillpath;
\draw[gp path] (4.696,0.985)--(4.696,1.338)--(5.127,1.338)--(5.127,0.985)--cycle;
\def\gpfillpath{(5.415,0.985)--(5.847,0.985)--(5.847,1.187)--(5.415,1.187)--cycle}
\gpfill{color=gpbgfillcolor} \gpfillpath;
\gpfill{rgb color={0.000,1.000,0.000},gp pattern 2,pattern color=.} \gpfillpath;
\draw[gp path] (5.415,0.985)--(5.415,1.186)--(5.846,1.186)--(5.846,0.985)--cycle;
\def\gpfillpath{(6.133,0.985)--(6.565,0.985)--(6.565,1.211)--(6.133,1.211)--cycle}
\gpfill{color=gpbgfillcolor} \gpfillpath;
\gpfill{rgb color={0.000,1.000,0.000},gp pattern 2,pattern color=.} \gpfillpath;
\draw[gp path] (6.133,0.985)--(6.133,1.210)--(6.564,1.210)--(6.564,0.985)--cycle;
\gpcolor{color=gp lt color border}
\node[gp node right] at (6.035,3.676) {Metadata updates + Messages};
\def\gpfillpath{(6.219,3.599)--(7.135,3.599)--(7.135,3.753)--(6.219,3.753)--cycle}
\gpfill{color=gpbgfillcolor} \gpfillpath;
\gpfill{rgb color={0.000,0.000,1.000},gp pattern 3,pattern color=.} \gpfillpath;
\gpcolor{rgb color={0.000,0.000,1.000}}
\draw[gp path] (6.219,3.599)--(7.135,3.599)--(7.135,3.753)--(6.219,3.753)--cycle;
\def\gpfillpath{(1.823,1.195)--(2.255,1.195)--(2.255,1.584)--(1.823,1.584)--cycle}
\gpfill{color=gpbgfillcolor} \gpfillpath;
\gpfill{rgb color={0.000,0.000,1.000},gp pattern 3,pattern color=.} \gpfillpath;
\draw[gp path] (1.823,1.195)--(1.823,1.583)--(2.254,1.583)--(2.254,1.195)--cycle;
\def\gpfillpath{(2.541,1.200)--(2.973,1.200)--(2.973,1.550)--(2.541,1.550)--cycle}
\gpfill{color=gpbgfillcolor} \gpfillpath;
\gpfill{rgb color={0.000,0.000,1.000},gp pattern 3,pattern color=.} \gpfillpath;
\draw[gp path] (2.541,1.200)--(2.541,1.549)--(2.972,1.549)--(2.972,1.200)--cycle;
\def\gpfillpath{(3.260,1.284)--(3.692,1.284)--(3.692,1.724)--(3.260,1.724)--cycle}
\gpfill{color=gpbgfillcolor} \gpfillpath;
\gpfill{rgb color={0.000,0.000,1.000},gp pattern 3,pattern color=.} \gpfillpath;
\draw[gp path] (3.260,1.284)--(3.260,1.723)--(3.691,1.723)--(3.691,1.284)--cycle;
\def\gpfillpath{(3.978,1.225)--(4.410,1.225)--(4.410,1.776)--(3.978,1.776)--cycle}
\gpfill{color=gpbgfillcolor} \gpfillpath;
\gpfill{rgb color={0.000,0.000,1.000},gp pattern 3,pattern color=.} \gpfillpath;
\draw[gp path] (3.978,1.225)--(3.978,1.775)--(4.409,1.775)--(4.409,1.225)--cycle;
\def\gpfillpath{(4.696,1.338)--(5.128,1.338)--(5.128,1.757)--(4.696,1.757)--cycle}
\gpfill{color=gpbgfillcolor} \gpfillpath;
\gpfill{rgb color={0.000,0.000,1.000},gp pattern 3,pattern color=.} \gpfillpath;
\draw[gp path] (4.696,1.338)--(4.696,1.756)--(5.127,1.756)--(5.127,1.338)--cycle;
\def\gpfillpath{(5.415,1.186)--(5.847,1.186)--(5.847,1.914)--(5.415,1.914)--cycle}
\gpfill{color=gpbgfillcolor} \gpfillpath;
\gpfill{rgb color={0.000,0.000,1.000},gp pattern 3,pattern color=.} \gpfillpath;
\draw[gp path] (5.415,1.186)--(5.415,1.913)--(5.846,1.913)--(5.846,1.186)--cycle;
\def\gpfillpath{(6.133,1.210)--(6.565,1.210)--(6.565,1.884)--(6.133,1.884)--cycle}
\gpfill{color=gpbgfillcolor} \gpfillpath;
\gpfill{rgb color={0.000,0.000,1.000},gp pattern 3,pattern color=.} \gpfillpath;
\draw[gp path] (6.133,1.210)--(6.133,1.883)--(6.564,1.883)--(6.564,1.210)--cycle;
\gpcolor{color=gp lt color border}
\draw[gp path] (1.320,3.393)--(1.320,0.985)--(7.067,0.985)--(7.067,3.393)--cycle;
\gpdefrectangularnode{gp plot 1}{\pgfpoint{1.320cm}{0.985cm}}{\pgfpoint{7.067cm}{3.393cm}}
\end{tikzpicture}

%% file: charts/louvain-naive.tex
\begin{tikzpicture}[gnuplot]
\path (0.000,0.000) rectangle (7.620,4.318);
\gpcolor{color=gp lt color border}
\gpsetlinetype{gp lt border}
\gpsetlinewidth{1.00}
\draw[gp path] (1.320,0.985)--(1.500,0.985);
\draw[gp path] (7.067,0.985)--(6.887,0.985);
\node[gp node right] at (1.136,0.985) {$0$};
\draw[gp path] (1.320,1.405)--(1.500,1.405);
\draw[gp path] (7.067,1.405)--(6.887,1.405);
\node[gp node right] at (1.136,1.405) {$20$};
\draw[gp path] (1.320,1.825)--(1.500,1.825);
\draw[gp path] (7.067,1.825)--(6.887,1.825);
\node[gp node right] at (1.136,1.825) {$40$};
\draw[gp path] (1.320,2.245)--(1.500,2.245);
\draw[gp path] (7.067,2.245)--(6.887,2.245);
\node[gp node right] at (1.136,2.245) {$60$};
\draw[gp path] (1.320,2.665)--(1.500,2.665);
\draw[gp path] (7.067,2.665)--(6.887,2.665);
\node[gp node right] at (1.136,2.665) {$80$};
\draw[gp path] (1.320,3.085)--(1.500,3.085);
\draw[gp path] (7.067,3.085)--(6.887,3.085);
\node[gp node right] at (1.136,3.085) {$100$};
\draw[gp path] (2.038,0.985)--(2.038,1.165);
\draw[gp path] (2.038,3.085)--(2.038,2.905);
\node[gp node center] at (2.038,0.677) {1};
\draw[gp path] (2.757,0.985)--(2.757,1.165);
\draw[gp path] (2.757,3.085)--(2.757,2.905);
\node[gp node center] at (2.757,0.677) {2};
\draw[gp path] (3.475,0.985)--(3.475,1.165);
\draw[gp path] (3.475,3.085)--(3.475,2.905);
\node[gp node center] at (3.475,0.677) {3};
\draw[gp path] (4.194,0.985)--(4.194,1.165);
\draw[gp path] (4.194,3.085)--(4.194,2.905);
\node[gp node center] at (4.194,0.677) {4};
\draw[gp path] (4.912,0.985)--(4.912,1.165);
\draw[gp path] (4.912,3.085)--(4.912,2.905);
\node[gp node center] at (4.912,0.677) {5};
\draw[gp path] (5.630,0.985)--(5.630,1.165);
\draw[gp path] (5.630,3.085)--(5.630,2.905);
\node[gp node center] at (5.630,0.677) {6};
\draw[gp path] (6.349,0.985)--(6.349,1.165);
\draw[gp path] (6.349,3.085)--(6.349,2.905);
\node[gp node center] at (6.349,0.677) {7};
\draw[gp path] (1.320,3.085)--(1.320,0.985)--(7.067,0.985)--(7.067,3.085)--cycle;
\node[gp node center,rotate=-270] at (0.276,2.035) {Runtime (sec)};
\node[gp node center] at (4.193,0.215) {Louvain level};
\node[gp node right] at (5.391,3.984) {Compute};
\def\gpfillpath{(5.575,3.907)--(6.491,3.907)--(6.491,4.061)--(5.575,4.061)--cycle}
\gpfill{color=gpbgfillcolor} \gpfillpath;
\gpfill{rgb color={0.000,1.000,0.000},gp pattern 2,pattern color=.} \gpfillpath;
\gpcolor{rgb color={0.000,1.000,0.000}}
\draw[gp path] (5.575,3.907)--(6.491,3.907)--(6.491,4.061)--(5.575,4.061)--cycle;
\def\gpfillpath{(1.823,0.985)--(2.255,0.985)--(2.255,1.098)--(1.823,1.098)--cycle}
\gpfill{color=gpbgfillcolor} \gpfillpath;
\gpfill{rgb color={0.000,1.000,0.000},gp pattern 2,pattern color=.} \gpfillpath;
\draw[gp path] (1.823,0.985)--(1.823,1.097)--(2.254,1.097)--(2.254,0.985)--cycle;
\def\gpfillpath{(2.541,0.985)--(2.973,0.985)--(2.973,1.094)--(2.541,1.094)--cycle}
\gpfill{color=gpbgfillcolor} \gpfillpath;
\gpfill{rgb color={0.000,1.000,0.000},gp pattern 2,pattern color=.} \gpfillpath;
\draw[gp path] (2.541,0.985)--(2.541,1.093)--(2.972,1.093)--(2.972,0.985)--cycle;
\def\gpfillpath{(3.260,0.985)--(3.692,0.985)--(3.692,1.098)--(3.260,1.098)--cycle}
\gpfill{color=gpbgfillcolor} \gpfillpath;
\gpfill{rgb color={0.000,1.000,0.000},gp pattern 2,pattern color=.} \gpfillpath;
\draw[gp path] (3.260,0.985)--(3.260,1.097)--(3.691,1.097)--(3.691,0.985)--cycle;
\def\gpfillpath{(3.978,0.985)--(4.410,0.985)--(4.410,1.050)--(3.978,1.050)--cycle}
\gpfill{color=gpbgfillcolor} \gpfillpath;
\gpfill{rgb color={0.000,1.000,0.000},gp pattern 2,pattern color=.} \gpfillpath;
\draw[gp path] (3.978,0.985)--(3.978,1.049)--(4.409,1.049)--(4.409,0.985)--cycle;
\def\gpfillpath{(4.696,0.985)--(5.128,0.985)--(5.128,1.072)--(4.696,1.072)--cycle}
\gpfill{color=gpbgfillcolor} \gpfillpath;
\gpfill{rgb color={0.000,1.000,0.000},gp pattern 2,pattern color=.} \gpfillpath;
\draw[gp path] (4.696,0.985)--(4.696,1.071)--(5.127,1.071)--(5.127,0.985)--cycle;
\def\gpfillpath{(5.415,0.985)--(5.847,0.985)--(5.847,1.064)--(5.415,1.064)--cycle}
\gpfill{color=gpbgfillcolor} \gpfillpath;
\gpfill{rgb color={0.000,1.000,0.000},gp pattern 2,pattern color=.} \gpfillpath;
\draw[gp path] (5.415,0.985)--(5.415,1.063)--(5.846,1.063)--(5.846,0.985)--cycle;
\def\gpfillpath{(6.133,0.985)--(6.565,0.985)--(6.565,1.098)--(6.133,1.098)--cycle}
\gpfill{color=gpbgfillcolor} \gpfillpath;
\gpfill{rgb color={0.000,1.000,0.000},gp pattern 2,pattern color=.} \gpfillpath;
\draw[gp path] (6.133,0.985)--(6.133,1.097)--(6.564,1.097)--(6.564,0.985)--cycle;
\gpcolor{color=gp lt color border}
\node[gp node right] at (5.391,3.676) {Reconstruction + I/O};
\def\gpfillpath{(5.575,3.599)--(6.491,3.599)--(6.491,3.753)--(5.575,3.753)--cycle}
\gpfill{color=gpbgfillcolor} \gpfillpath;
\gpfill{rgb color={1.000,0.000,0.000},gp pattern 3,pattern color=.} \gpfillpath;
\gpcolor{rgb color={1.000,0.000,0.000}}
\draw[gp path] (5.575,3.599)--(6.491,3.599)--(6.491,3.753)--(5.575,3.753)--cycle;
\def\gpfillpath{(1.823,1.097)--(2.255,1.097)--(2.255,2.704)--(1.823,2.704)--cycle}
\gpfill{color=gpbgfillcolor} \gpfillpath;
\gpfill{rgb color={1.000,0.000,0.000},gp pattern 3,pattern color=.} \gpfillpath;
\draw[gp path] (1.823,1.097)--(1.823,2.703)--(2.254,2.703)--(2.254,1.097)--cycle;
\def\gpfillpath{(2.541,1.093)--(2.973,1.093)--(2.973,2.359)--(2.541,2.359)--cycle}
\gpfill{color=gpbgfillcolor} \gpfillpath;
\gpfill{rgb color={1.000,0.000,0.000},gp pattern 3,pattern color=.} \gpfillpath;
\draw[gp path] (2.541,1.093)--(2.541,2.358)--(2.972,2.358)--(2.972,1.093)--cycle;
\def\gpfillpath{(3.260,1.097)--(3.692,1.097)--(3.692,2.123)--(3.260,2.123)--cycle}
\gpfill{color=gpbgfillcolor} \gpfillpath;
\gpfill{rgb color={1.000,0.000,0.000},gp pattern 3,pattern color=.} \gpfillpath;
\draw[gp path] (3.260,1.097)--(3.260,2.122)--(3.691,2.122)--(3.691,1.097)--cycle;
\def\gpfillpath{(3.978,1.049)--(4.410,1.049)--(4.410,1.923)--(3.978,1.923)--cycle}
\gpfill{color=gpbgfillcolor} \gpfillpath;
\gpfill{rgb color={1.000,0.000,0.000},gp pattern 3,pattern color=.} \gpfillpath;
\draw[gp path] (3.978,1.049)--(3.978,1.922)--(4.409,1.922)--(4.409,1.049)--cycle;
\def\gpfillpath{(4.696,1.071)--(5.128,1.071)--(5.128,1.975)--(4.696,1.975)--cycle}
\gpfill{color=gpbgfillcolor} \gpfillpath;
\gpfill{rgb color={1.000,0.000,0.000},gp pattern 3,pattern color=.} \gpfillpath;
\draw[gp path] (4.696,1.071)--(4.696,1.974)--(5.127,1.974)--(5.127,1.071)--cycle;
\def\gpfillpath{(5.415,1.063)--(5.847,1.063)--(5.847,1.792)--(5.415,1.792)--cycle}
\gpfill{color=gpbgfillcolor} \gpfillpath;
\gpfill{rgb color={1.000,0.000,0.000},gp pattern 3,pattern color=.} \gpfillpath;
\draw[gp path] (5.415,1.063)--(5.415,1.791)--(5.846,1.791)--(5.846,1.063)--cycle;
\def\gpfillpath{(6.133,1.097)--(6.565,1.097)--(6.565,1.590)--(6.133,1.590)--cycle}
\gpfill{color=gpbgfillcolor} \gpfillpath;
\gpfill{rgb color={1.000,0.000,0.000},gp pattern 3,pattern color=.} \gpfillpath;
\draw[gp path] (6.133,1.097)--(6.133,1.589)--(6.564,1.589)--(6.564,1.097)--cycle;
\gpcolor{color=gp lt color border}
\node[gp node right] at (5.391,3.368) {SAFS ingest};
\def\gpfillpath{(5.575,3.291)--(6.491,3.291)--(6.491,3.445)--(5.575,3.445)--cycle}
\gpfill{color=gpbgfillcolor} \gpfillpath;
\gpfill{rgb color={1.000,0.647,0.000},gp pattern 5,pattern color=.} \gpfillpath;
\gpcolor{rgb color={1.000,0.647,0.000}}
\draw[gp path] (5.575,3.291)--(6.491,3.291)--(6.491,3.445)--(5.575,3.445)--cycle;
\def\gpfillpath{(1.823,2.703)--(2.255,2.703)--(2.255,2.989)--(1.823,2.989)--cycle}
\gpfill{color=gpbgfillcolor} \gpfillpath;
\gpfill{rgb color={1.000,0.647,0.000},gp pattern 5,pattern color=.} \gpfillpath;
\draw[gp path] (1.823,2.703)--(1.823,2.988)--(2.254,2.988)--(2.254,2.703)--cycle;
\def\gpfillpath{(2.541,2.358)--(2.973,2.358)--(2.973,2.613)--(2.541,2.613)--cycle}
\gpfill{color=gpbgfillcolor} \gpfillpath;
\gpfill{rgb color={1.000,0.647,0.000},gp pattern 5,pattern color=.} \gpfillpath;
\draw[gp path] (2.541,2.358)--(2.541,2.612)--(2.972,2.612)--(2.972,2.358)--cycle;
\def\gpfillpath{(3.260,2.122)--(3.692,2.122)--(3.692,2.391)--(3.260,2.391)--cycle}
\gpfill{color=gpbgfillcolor} \gpfillpath;
\gpfill{rgb color={1.000,0.647,0.000},gp pattern 5,pattern color=.} \gpfillpath;
\draw[gp path] (3.260,2.122)--(3.260,2.390)--(3.691,2.390)--(3.691,2.122)--cycle;
\def\gpfillpath{(3.978,1.922)--(4.410,1.922)--(4.410,2.166)--(3.978,2.166)--cycle}
\gpfill{color=gpbgfillcolor} \gpfillpath;
\gpfill{rgb color={1.000,0.647,0.000},gp pattern 5,pattern color=.} \gpfillpath;
\draw[gp path] (3.978,1.922)--(3.978,2.165)--(4.409,2.165)--(4.409,1.922)--cycle;
\def\gpfillpath{(4.696,1.974)--(5.128,1.974)--(5.128,2.246)--(4.696,2.246)--cycle}
\gpfill{color=gpbgfillcolor} \gpfillpath;
\gpfill{rgb color={1.000,0.647,0.000},gp pattern 5,pattern color=.} \gpfillpath;
\draw[gp path] (4.696,1.974)--(4.696,2.245)--(5.127,2.245)--(5.127,1.974)--cycle;
\def\gpfillpath{(5.415,1.791)--(5.847,1.791)--(5.847,2.065)--(5.415,2.065)--cycle}
\gpfill{color=gpbgfillcolor} \gpfillpath;
\gpfill{rgb color={1.000,0.647,0.000},gp pattern 5,pattern color=.} \gpfillpath;
\draw[gp path] (5.415,1.791)--(5.415,2.064)--(5.846,2.064)--(5.846,1.791)--cycle;
\def\gpfillpath{(6.133,1.589)--(6.565,1.589)--(6.565,1.830)--(6.133,1.830)--cycle}
\gpfill{color=gpbgfillcolor} \gpfillpath;
\gpfill{rgb color={1.000,0.647,0.000},gp pattern 5,pattern color=.} \gpfillpath;
\draw[gp path] (6.133,1.589)--(6.133,1.829)--(6.564,1.829)--(6.564,1.589)--cycle;
\gpcolor{color=gp lt color border}
\draw[gp path] (1.320,3.085)--(1.320,0.985)--(7.067,0.985)--(7.067,3.085)--cycle;
\gpdefrectangularnode{gp plot 1}{\pgfpoint{1.320cm}{0.985cm}}{\pgfpoint{7.067cm}{3.085cm}}
\end{tikzpicture}

%% file: software.tex
\section{Software}

\graphyti{} is an open source library available through Python's \texttt{pip}
package manager under the name \texttt{graphyti}. To extend the library,
developers can visit \\
\href{https://github.com/flashxio/Graphyti}{https://github.com/flashxio/graphyti}.
Furthermore, we provide Docker integration for developers to reduce the barrier
to entry.

%% file: conclusions.tex
\section{Conclusions} \label{sec:gt:conclusions}

We present key principles that are critical to state-of-the-art
performance for vertex-centric, semi-external-memory
graph algorithms. Example applications within \graphyti{} illustrate
the positive performance effects of adopting these principles.
Finally, we improve the accessibility of SEM graph applications for users by
providing a high-level Python interface to \graphyti{}. 

%% file: author-bios.tex
\hfill \break
\vspace{-10pt}
\begin{wrapfigure}{l}{20mm}
    \includegraphics[width=0.75in,height=1in,clip,keepaspectratio]{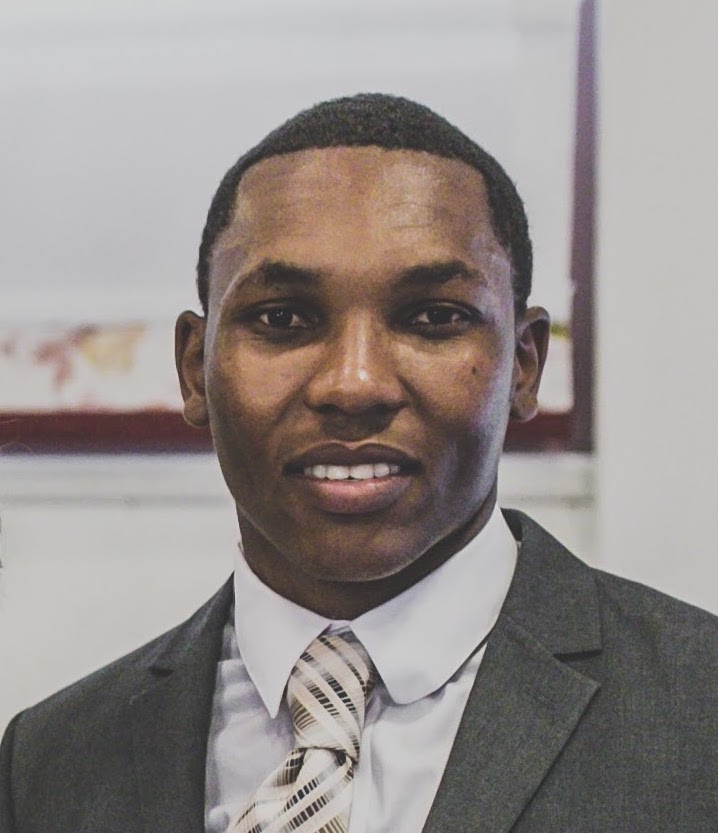}
\end{wrapfigure}\par
  \input{bios/disa}
  \par

\hfill \break
\vspace{-10pt}
\begin{wrapfigure}{l}{20mm}
    \includegraphics[width=0.75in,height=1in,clip,keepaspectratio]{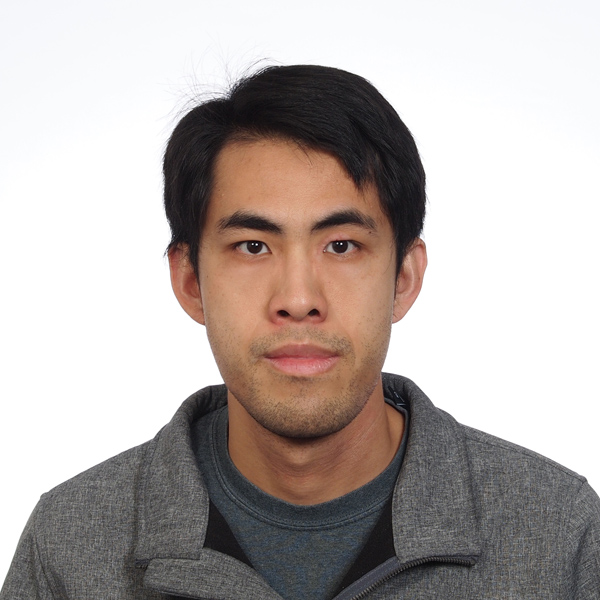}
\end{wrapfigure}\par
  \input{bios/da}
  \par

\hfill \break
\vspace{-10pt}
\begin{wrapfigure}{l}{20mm}
    \includegraphics[width=0.75in,height=1in,clip,keepaspectratio]{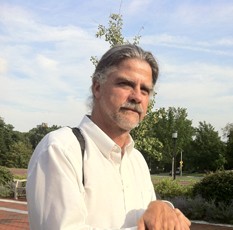}
\end{wrapfigure}\par
  \input{bios/cep}
  \par

\hfill \break
\vspace{-10pt}
\begin{wrapfigure}{l}{20mm}
    \includegraphics[width=0.75in,height=1in,clip,keepaspectratio]{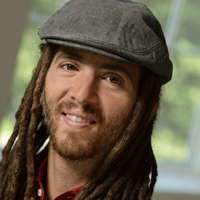}
\end{wrapfigure}\par
  \input{bios/jovo}
  \par

\hfill \break
\vspace{-10pt}
\begin{wrapfigure}{l}{20mm}
    \includegraphics[width=0.75in,height=1in,clip,keepaspectratio]{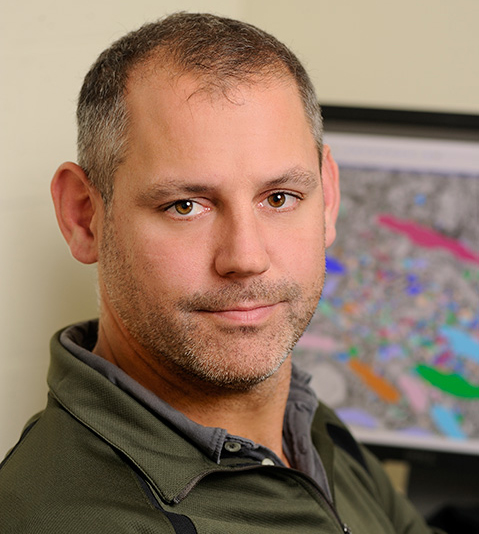}
\end{wrapfigure}\par
  \input{bios/randal}
  \par

%% file: bios/disa.tex
\textbf{Disa Mhembere} received his PhD in computer science from the
Johns Hopkins University in 2019.  His research interests lie in scalable framework
development for scientific computing and machine learning.

%% file: bios/da.tex
\textbf{Da Zheng} is an applied scientist at Amazon.
He received his PhD in computer science from the
Johns Hopkins University in 2016. His research interests are developing
frameworks for large-scale machine learning.

%% file: bios/cep.tex
\textbf{Carey E. Priebe} received a PhD in information technology
(computational statistics) from George Mason University in 1993.
Since 1994 he has been a professor in the Department of Applied
Mathematics and Statistics at Johns Hopkins University.

%% file: bios/jovo.tex
\textbf{Joshua T. Vogelstein} is an Assistant Professor in the Department of
Biomedical Engineering and the Institute for Computational Medicine at
Johns Hopkins University. He received his PhD in neuroscience from Johns
Hopkins University in 2009.

%% file: bios/randal.tex
\textbf{Randal Burns} earned his PhD in computer science in 2000 from
he U.C. Santa Cruz. He is a Professor and Chair in the
Computer Science Department at Johns Hopkins University.
His research interests lie in building the high-performance,
scalable data systems.